\documentclass[reprint,amsmath,amssymb,aps, pra]{revtex4-2}

\usepackage[linktocpage=true, colorlinks=true, allcolors=blue,]{hyperref}

\usepackage{tabularx} 
\usepackage{booktabs} 
\usepackage{graphicx}
\usepackage{physics} 
\usepackage{xcolor}
\usepackage[thinc]{esdiff} 
\usepackage{mathtools} 

\usepackage{amsthm}
\usepackage{subcaption} 

\newcommand{\defeq}{\vcentcolon=}
\newcommand{\eqdef}{=\vcentcolon}
\newcommand{\I}{\operatorname{I}}

\newcommand{\X}{\operatorname{X}}

\newcommand{\Y}{\operatorname{Y}}
\newcommand{\Z}{\operatorname{Z}}

\newcommand{\CCZ}{\operatorname{CCZ}}

\newcommand{\WX}{\operatorname{WX}}
\newcommand{\WZ}{\operatorname{WZ}}

\DeclareMathOperator*{\argmin}{argmin}

\newcommand{\R}{\mathbb{R}}
\newcommand{\C}{\mathbb{C}}

\newcommand{\Od}{\mathbb{O}}
\newcommand{\SE}[1]{\operatorname{SE}(#1)}
\newcommand{\SO}[1]{\operatorname{SO}(#1)}
\newcommand{\SU}[1]{\operatorname{SU}(#1)}
\newcommand{\su}[1]{\mathfrak{su}(#1)}
\newcommand{\F}{\mathcal{F}}
\newcommand{\G}{\mathcal{G}}
\newcommand{\g}{\mathfrak{g}}
\newcommand{\Xvec}{\mathfrak{X}}

\newcommand{\hess}{\operatorname{Hess}}
\newcommand{\Ad}{\operatorname{Ad}}
\newcommand{\ad}{\operatorname{ad}}
\newcommand{\dlog}{\operatorname{Dlog}}

\newcommand{\dexp}{\operatorname{dexp}}

\newcommand{\D}{\operatorname{D}}
\newcommand{\at}[2][]{#1|_{#2}}
\newcommand{\of}[2][]{#1\bigl({#2}\bigr)}
\newcommand{\TL}[2]{T_{#1}L_{#2}} 

\begin{document}
    \title{Iterative linear quadratic regulator on SU(N) for multi-qubit gate synthesis}
	
	\author{Dirk Heimann}
	\email[Contact author: ]{dirk.heimann@uni-bremen.de}
	\affiliation{University of Bremen, 28359 Bremen, Germany}
	\affiliation{Bremen University of Applied Science, 28199 Bremen, Germany}
	\author{Felix Wiebe}
	\affiliation{German Research Center for Artificial Intelligence (DFKI), 28359 Bremen, Germany}
	\author{Elie Mounzer}
	\affiliation{German Research Center for Artificial Intelligence (DFKI), 28359 Bremen, Germany}
	\author{Shivesh Kumar}
	\affiliation{Chalmers University of Technology, 412 96 Gothenburg, Sweden}
	
	\date{04.08.2026}
	

\begin{abstract}
	In quantum optimal control theory, gradient-based trajectory optimization techniques have proven versatile in designing multi-qubit quantum gates. Furthermore, incorporating the underlying Lie-group structure can accelerate the optimization process. In this work, we adapt the Lie-group formulation of the iterative linear quadratic regulator (iLQR) to the special unitary group SU(N) and apply it to quantum gate synthesis, systematically comparing it against the standard Euclidean iLQR formulation across multiple two- to five-qubit gates. We find that in the idealized, unconstrained setting, where all Lie-algebra basis elements are available as drive Hamiltonian terms, the Lie-group formulation converges faster than the Euclidean iLQR formulation. If drive terms are constrained to 2-local Hamiltonian terms, the Lie-group variant converges faster in early optimization iterations, but exhibits greater sensitivity to initialization and a stronger tendency towards local minima. These results demonstrate that incorporating Lie-group geometry into iLQR substantially improves convergence and highlight important next steps for improvements in constrained control settings.
\end{abstract}

	
	\maketitle
	
	\section{Introduction}

Quantum optimal control (QOC) aims to design electromagnetic control fields that steer quantum systems to perform desired operations with high fidelity~\cite{koch2022}. A prominent application is the synthesis of control pulses to implement quantum gates on hardware platforms such as superconducting circuits~\cite[Sec.~5]{koch2022}. While physically motivated analytic solutions for pulse schemes exist for specific Hamiltonians and gates, they do not generalize systematically, motivating the use of numerical methods. Among many others, common numerical QOC methods include Gradient ascent pulse engineering (GRAPE)~\cite{khaneja2005}, second-order projection-operator Newton methods~\cite{shao2022}, direct collocation~\cite{trowbridge2023}, and trajectory optimization techniques such as the iterative linear quadratic regulator (iLQR)~\cite{heimann2025ilqr} and its augmented Lagrangian version for robust control~\cite{propson2022} based on ALTRO~\cite{howell2019}.
Despite their generality, these optimal control methods can become computationally demanding when applied to pulse-level quantum gate design: they often require many iterations, especially when the control is highly discretized in time, and incur additional overhead when second-order gradients are used because they require evaluating the Hessian.

Since quantum dynamics evolve on the special unitary Lie group $\SU{N}$, explicitly incorporating the underlying group structure offers an opportunity to improve convergence while only moderately increasing the computational effort.
Early work in this direction studied geodesics on $\SU{N}$ to establish upper and lower bounds on the number of gates required to synthesize unitary operations exactly~\cite{nielson2006} and generalized these results to the optimal control cost~\cite{nielson2006geom}. These bounds were later refined using Suzuki–Trotter decompositions~\cite{bhattacharyya2020}.
From a numerical perspective, incorporating analytic derivative information into second-order GRAPE methods has been shown to yield substantial wall-clock speedups via the auxiliary matrix method~\cite{goodwin2016} and ESCALADE-based approaches~\cite{foroozandeh2021, goodwin2023}.
While analytic expressions for matrix exponentials required by ESCALADE are known for $N\in \{2,3,4\}$~\cite{kaiser2022}, obtaining analytical formulas for the corresponding directional propagator derivatives becomes increasingly challenging for higher-dimensional systems relevant to multi-qubit superconducting circuits.
Therefore, for multi-qubit quantum gate synthesis, GRAPE has been extended to exploit geodesics toward the target unitary in the geodesic pulse engineering (GEOPE) method, for both time-independent~\cite{lewis2024} and time-dependent Hamiltonians~\cite{lewis2025}, achieving high fidelities with fewer optimization steps than GRAPE.

The Euclidean iLQR method and its second-order gradient version, differential dynamic programming (DDP)~\cite{jacobson1970}, can be extended to incorporate Lie-group structure by reformulating the dynamics on Lie algebra elements and cost functions defined on Lie group elements~\cite{boutselis2021}, as well as for a computationally simpler Lie-algebraic cost function~\cite{teng2022}. Both works validate their algorithms numerically on the dynamics of a rigid satellite evolving on $\SO{3}\times \R^3$. A similar extension to LQR-based controls on the Lie group $\SE{3}$ has been proposed~\cite{2023_icra_tvlqr_SE3}.

In this work, we extend the state of the art by providing necessary adaptations to the Lie-group iLQR formulation for quantum gate synthesis on $\SU{N}$, presenting a worst-case time-complexity analysis and systematically evaluating its numerical behavior. In particular, we define Lie-algebraic states consistent with the definitions for quantum optimal control~\cite{lewis2024} to adapt iLQR to a Lie-theoretic setting by determining analytic expressions for the first-order dynamic approximation via the adjoint representation of $\su{N}$ and $\SU{N}$. Furthermore, we show that the proposed Lie-group iLQR method aligns well with the smoothing effect for pulse amplitudes~\cite{heimann2025ilqr} and introduce and test numerically the restriction of accessible control Hamiltonian terms, which is essential for quantum control on hardware.
Specifically, our contributions can be summarized as follows:

\begin{itemize}
	\item We show how to adapt iLQR on $\SU{N}$ for quantum gate synthesis, demonstrate how the method can naturally incorporate smoothed controls, and provide a worst-case time-complexity analysis.
	\item We provide numerical comparisons between the Lie-group iLQR and its Euclidean formulation for the full algebra control setting, showing that the Lie-group formulation achieves infidelities up to machine precision in a single optimization iteration across all tested gates in this idealized setting.
	\item We present numerical results for 2-local restricted controls, which show that the Lie-group formulation achieves lower infidelities at early optimization steps but has increased sensitivity to initialization, a tendency toward local minima, and, similar to other methods, degrading performance for larger qubit systems.
\end{itemize}

The remainder of this paper is organized as follows. Section~\ref{sec:prelim} introduces necessary Lie-theoretic definitions and notation. Section~\ref{sec:method} extends iLQR to Lie groups by incorporating the necessary adaptations for $\SU{N}$ and defines the Lie-algebraic equations of motion and cost functions for $\SU{N}$. The natural extension to smooth controls and a worst-case time-complexity analysis are also provided. Section~\ref{sec:results} presents numerical simulation results for the optimization of two- to five-qubit gates. Finally, Section~\ref{sec:discussion} discusses the results and highlights limitations that identify future work.
	
	\section{Preliminaries and notation}
    \label{sec:prelim}

Unitary matrices $U$ with determinant equal to one represent noise-free operations on quantum states and form the Lie group $\SU{N}$.
Its associated Lie algebra $\su{N}$ consists of skew-Hermitian, traceless matrices and is closed under the commutator
$[\eta^\wedge, \xi^\wedge] = \eta^\wedge \xi^\wedge - \xi^\wedge \eta^\wedge$.
\begin{align*}
    \SU{N} &= \{ U \in \C^{N\times N} | U^{-1} = U^\dagger, \det(U) = 1\} \\
    \su{N} &= \{ \eta^\wedge \in \C^{N\times N} |
\eta^\wedge = -\eta^{\wedge\dagger}, \tr(\eta^\wedge)= 0\}.
\end{align*}
Both are differentiable manifolds of dimension $N^2-1$
where $N = l^q$ for a system of $q$ qubits with $l$ levels each.
We denote elements of a general Lie algebras by $\eta^\wedge, \xi^\wedge \in \g$
with corresponding vector representations $\eta, \xi \in \R^{\dim(\g)}$ and group elements by $g\in G$.

An isomorphism between Lie algebra and real vector space is introduced via the wedge-  and its inverse map
\begin{align*}
    (\cdot)^\wedge \colon \mathbb{R}^{\dim(\g)} \to \g \quad &\text{and} \quad 
    (\cdot)^\vee \colon \g \to \mathbb{R}^{\dim(\g)}
\end{align*}
such that $(\xi^{\wedge})^{\vee} = \xi$ for all $\xi\in\R^{\dim(\g)}$.

The adjoint representation of the group, $\Ad_g$, and the algebra, $\ad_{\eta^\wedge}$, are defined as 
\begin{align*}
\Ad_g \colon \g \to \g \quad &\text{with} \quad \Ad_g(\xi^\wedge) = g\xi^\wedge g^{-1} \\ 
\ad_{\eta^\wedge} \colon \g \to \g \quad &\text{with} \quad \ad_{\eta^\wedge}(\xi^\wedge) = [\eta^\wedge, \xi^\wedge].
\end{align*}

The exponential map $\exp \colon \g \to \G$ is a local diffeomorphism defined by the series $\exp(\xi^\wedge) = \sum_{k=0}^\infty \frac{1}{k!}(\xi^\wedge)^k$. The differential of the exponential map $\dexp \colon \g \times \g \to \g$ is defined by the right-trivialized tangent of the exponential map at $\eta^\wedge$ in direction $\xi^\wedge$ and given by the series~\cite{iserles2000}
\begin{align}
	\label{eq:dexp_series}
	\dexp_{\eta^\wedge}(\xi^\wedge) 
	\nonumber
	&= \xi^\wedge + \frac{1}{2!}[\eta^\wedge, \xi^\wedge] + \frac{1}{3!}[\eta^\wedge, [\eta^\wedge,\xi^\wedge]] + \cdots \\
	&=\sum_{j=0}^{\infty} \frac{1}{(j+1)!} \ad_{\eta^\wedge}^j({\xi^\wedge}) \text{ and} \\
	\dexp^{-1}_{\eta^\wedge}(\xi^\wedge) \nonumber
	&= \xi^\wedge - \frac{1}{2}[\eta^\wedge, \xi^\wedge] + \frac{1}{12}[\eta^\wedge, [\eta^\wedge,\xi^\wedge]] + \cdots \\
	\label{eq:dexp_inv_series}
	&=\sum_{j=0}^{\infty} \frac{B^-_j}{j!} \ad_{\eta^\wedge}^j({\xi^\wedge})
\end{align}
with Bernoulli numbers $B^-_j = \sum_{m=0}^j\frac{1}{m+1}\sum_{n=0}^m \binom{m}{n}(-1)^n n^j$. Further Lie group definitions and basics are provided in App.~\ref{app:group_ilqr_basics}.
	
	\section{Method}
	\label{sec:method}

\subsection{Iterative linear quadratic regulator on Lie groups}
    Trajectory optimization aims to solve the following discrete-time, finite-horizon optimal control problem	
	\begin{align*}
		\min_{u_{0:T-1}} &J(g_{0:T}, u_{0:T-1}) \\
		J(g_{0:T}, u_{0:T-1}) &=  l_F(g_T) + \sum_{k=0}^{T-1} l(g_k, u_k)\\
		\text{s.t.} \quad g_{k+1} &= f(g_k, u_k), \quad g_0 = g_\text{init}. 
	\end{align*}
	The time is discretized by constant step size $\Delta t$ yielding the set $u_{0:T-1} = \{u_k\}_0^{T-1}$ of $T$-many controls $u_k\in\R^{N_u}$. The equation of motion $f\colon G\times \R^{N_u} \to G$ is once differentiable and propagates state $g_k$ to $g_{k+1}$ by applying controls $u_k$. The cost function $J$ is composed of a twice differentiable running $l_k\colon\G\times\R^{N_u} \to \R$ and final cost $l_F\colon \G \to \R$.
	
	The time-discrete action-value function is defined as $Q_k(g_k, u_k) = l(g_k,u_k) + V_{k+1}(f(g_{k}, u_k))$ where the value (cost-to-go) function is $V_k = V(g_k) = \min_{u_{k:T-1}} J(g_{k:T},u_{k:T-1})$.
    Bellman's principle of optimality states that, along an optimal trajectory, the remaining decisions from any intermediate state must themselves form an optimal policy~\cite{bellman1954}. This yields the recursive Bellman equation 
	\begin{align}\label{eq:cost-to-go}
		V_k\of{g_k} &=	\min_{u_k}\left[ l(g_k, u_k) + V_{k+1}(f(g_k, u_k))\right] \nonumber \\
		&= \min_{u_{k:K-1}} Q_k(g_k, u_k)
	\end{align}
	with $V_T = l_F(g_T)$.
    The solution to this equation is approximated locally by second-order approximations in Euclidean space in differential dynamic programming (DDP)~\cite{mayne1966}. Neglecting the second-order derivatives of the system dynamics yields the iterative linear quadratic regulator (iLQR) formulation~\cite{li2004}. For Lie-group elements $g_k\in\G$, locally optimal updates are obtained using 1st- and 2nd-order Taylor expansions on Lie groups~\cite{boutselis2021}.
    Therefore, we introduce perturbations of states $g_{\epsilon, k}$ and controls $u_{\epsilon, k}$ as
    \begin{align}
		\label{eq:su2_perturbed_group}
		g_{\epsilon, k}   &= g_k\exp(i\eta^\wedge_k) \\
		\label{eq:su2_perturbed_algebra}
		u_{\epsilon, k} &= u_k + \delta u_k
	\end{align}
	where $i\eta^\wedge\in \g$.
    We provide the details for Taylor expansions on Lie groups in App.~\ref{sec:app_taylor}, which enables us to express the perturbed action-value function $Q(g_{\epsilon,k}, u_{\epsilon,k})$ via Eq.~\eqref{eq:app_taylor_f_gu} up to second order as
	\begin{align*}
        \delta Q_k(g_{\epsilon,k},u_{\epsilon,k})
        \approx{}&
        Q_{g,k}(\eta_k^\wedge)
        +Q_{u,k}(\delta u_k)
        \nonumber\\
        &+\frac12 Q_{gg,k}(\eta_k^\wedge,\eta_k^\wedge)
        +\frac12 Q_{ug,k}(\delta u_k,\eta_k^\wedge)
        \nonumber\\
        &+\frac12 Q_{gu,k}(\eta_k^\wedge,\delta u_k)
        +\frac12 Q_{uu,k}(\delta u_k,\delta u_k)
    \end{align*}
    with individual terms
	\begin{align*}
		Q_k &= l_k + V_{k+1} \\
		Q_{g,k} &= l_{g,k} + V_{g,k+1} f_{g,k} \\
		Q_{u,k} &= l_{u,k} + V_{x,k+1} f_{u,k} \\
		Q_{gg,k} &= l_{gg,k} + f_{g,k}^T V_{gg,k+1} f_{g,k} \\ 
		Q_{uu,k} &= l_{uu,k} + f_{u,k}^T V_{gg,k+1} f_{u,k} \\
		Q_{ug,k} &= l_{ug,k} + f_{u,k}^T V_{ug,k+1} f_{g,k} \\
		Q_{gu,k} &= Q_{ug,k}^T.
	\end{align*}
	The required derivatives are obtained from corresponding perturbations of the value and cost functions. The change in the value function $V\of{g_{\epsilon, k}}$ with its first and second derivatives is derived from the perturbation of $g_k$
	\begin{align*}
		\of[\delta V_k]{g_{\epsilon, k}} &= V_{g,k} \of{\eta^\wedge_k} + \frac{1}{2}V_{gg,k}\of{\eta^\wedge_k,\eta^\wedge_k}
		+ \order{\norm{\eta^\wedge_k}^3} \\
		V_{g,k} &= \TL{e}{g_k}^* \circ \D_g V\of{g_k} \\
		V_{gg,k} &= \TL{e}{g_k}^* \circ \hess^{(0)}_gV\of{g_k} \circ \TL{e}{g_k} \\ 
		V_{g,T} &= \TL{e}{g_T}^* \circ \D_g l_f\of{g_T} \\
		V_{gg,T} &= \TL{e}{g_T}^* \circ \hess^{(0)}l_f\of{g_T} \circ \TL{e}{g_T}.
	\end{align*}
	Analogously, perturbing $g_k$ and $u_k$ determines the change in the loss function $l\of{g_{\epsilon, k}, u_{\epsilon, k}}$ with its first- and second-order derivative terms:
	\begin{align*}
		\delta \of[l_k]{g_{\epsilon, k}, u_{\epsilon, k}}
        \approx{}& \of[l_{g,k}]{\eta^\wedge_k} + \of[l_{u,k}]{\delta u_k} \\
		&+\frac{1}{2}\of[l_{gg,k}]{\eta^\wedge_k}\of{\eta^\wedge_k} +
		\frac{1}{2}\of[l_{u g,k}]{\delta u_k}\of{\eta^\wedge_k} \\
		&+\frac{1}{2}\of[l_{gu,k}]{\eta^\wedge_k}\of{\delta u_k} +
		\frac{1}{2}l_{uu,k}\of{\delta u_k}\of{\delta u_k} \\
		l_{g,k} &= \TL{e}{g_k}^* \circ \D_g l\of{g_k, u_k} \\
		l_{u,k} &= \D_u l\of{g_k, u_k} \\
		l_{gg,k} &= \TL{e}{g_k}^* \circ \hess_g^{(0)}l\of{g_k, u_k} \circ \TL{e}{g_k} \\
		l_{gu,k} &= \TL{e}{g_k}^* \circ \D_u\D_g l\of{g_k, u_k} \\
		l_{ug,k} &= \TL{e}{g_k}^* \circ \D_g\D_u l\of{g_k, u_k}.
	\end{align*}

	Taking the derivative to determine the minimum in Eq.~\eqref{eq:cost-to-go} with respect to $\delta u_k$ yields control updates $\delta u_k^\star$ that minimize the cost-to-go locally:
	\begin{align*}
		\delta u_k^\star &= \argmin_{\delta u_k}Q(g_{\epsilon,k}, u_{\epsilon,k}) \\
		&= -Q_{uu,k}^{-1} Q_{u,k}-Q_{uu,k}^{-1} Q_{ug,k}\eta_k\\
		&\eqdef \kappa_k + K_k \eta_k.
	\end{align*}

	The value function is then updated using the improved second-order update formulas~\cite{todorov2005, tassa2012}:
	\begin{align*}
		V_{g,k} &= Q_{g,k} + K_k^T Q_{uu,k}^{-1}\kappa_k + K_k^T Q_{u,k} + Q_{ux,k}^{T}\kappa_k \\
		V_{gg,k} &= Q_{gg,k} + K_k^T Q_{uu,k}^{-1}K_k + K_k^T Q_{ux,k} + Q_{ux,k}^{T}K_k.
	\end{align*}
	Starting from the terminal condition $V_T = l_f(g_T)$, a backward recursion for $k\in\{T-1, \ldots, 0\}$ is used to compute updates of the value function as well as the feedback gain $K_k$ and feed-forward term $\kappa_k$. The invertibility of $Q_{uu}$ is ensured by incorporating a Levenberg-Marquardt regularization scheme~\cite{mastalli2022}. During the forward pass, the feedback gains are applied to compute a new control sequence and the corresponding state trajectory, starting from the fixed start state $g_0^\text{new} = g_\text{init}$ and applying:
	\begin{align*}
		u^\text{new}_k &= u_k + \alpha \kappa_k + K_k \eta_k \\
		g^\text{new}_{k+1} &= f(g^\text{new}_k, u^\text{new}_k).
	\end{align*}
	A line search over the parameter $\alpha$, combined with the Goldstein acceptance criteria~\cite{mastalli2022}, prevents the update from moving too far away from the reference trajectory around which the system dynamics are locally linear. The backward and forward passes are alternated until convergence.
	
	In the following sections, we provide explicit expressions for the derivatives $f_{g, k}, f_{u, k}$ and $l_{g,k}$, $l_{u,k}$, $l_{gg,k}$, $l_{uu,k}$, $l_{gu,k}$ that are calculated within the backward pass.		

\subsection{Equations of motion in su(N)} \label{sec:eqm}
	The Schrödinger equation describes the time-discretized dynamics
	\begin{align}
		\label{eq:suN_eqm_group}
		U_{k+1} &= \exp(i(H_0 + H_du_k)) \sigma\Delta t) U_k \\ \nonumber
		&= \exp(\xi_k^\wedge \Delta t)U_k \\
		\nonumber 
		\Rightarrow \xi_k^\wedge &= \xi_k E \in \su{N}, \\ \nonumber
		\xi_k &= (H_0 + H_du_k) \in \R^{\dim(\su{N})}, \\
		\label{eq:su2_pauli_to_algebra_perturb}
		\delta \xi_k &= H_d \delta u_k
	\end{align}
	where $H_d\in \R$ denotes a scalar drive coefficient and $H_0, u^k\in \R^{\dim(SU(N)}$ denote drift and control vectors. Furthermore, $\{E_j\}$ denotes a basis of $\su{N}$ such that $\xi_k E = \sum_{j=1}^{\dim(\su{N})}\xi^j_k E_j$.
	
	For $N=2$, the basis elements are given by $\{E_j\} = \{i\sigma_j\}$, where $\sigma_i$ are the Pauli operators defined in Eq.~\eqref{eq:app_pauli_basis}. For larger $N$, calculating the Kronecker product of single-qubit Pauli operators forms a basis $i\mathcal{P}$ for $\su{N}$ by 
	\begin{align*}
		\mathcal{P} = \{ \sigma_{\alpha_1} \otimes \ldots \otimes \sigma_{\alpha_q} \} \setminus  \{\sigma_{\I} \otimes \ldots \otimes \sigma_{\I}\}
	\end{align*}
	for $\alpha_j \in \{\I,\X,\Y,\Z\}$~\cite{lewis2024}.

    By repeatedly applying the first equality in Eq.~\eqref{eq:suN_eqm_group} we get:
	\begin{align} \nonumber
		U_T &= U_{T-1} \ldots U_t \ldots U_1 U_\text{init} \\ \nonumber
		U_\text{init} U_T U_\text{init}^{-1} &= U_\text{init} U_{T-1} \ldots U_t \ldots U_1 \\ \nonumber
		U^\prime_T &= U_\text{init} U_{T-1} \ldots U_t \ldots U_1 \\
		\label{eq:eqm_rev}
		U_{k+1} &= U_{k} \exp(\xi^\wedge_k \Delta t) \text{ for } k = T -1 - t.
	\end{align}
	We initialize the propagation with $U_\text{init} = \sigma^{\otimes q}_{\I}$ which implies $U^\prime_T = U_T$. Therefore, the obtained controls $\xi_k$ have to be evaluated in reverse order, $t = T-1-k$, to ensure consistency with the Schrödinger equation.

    For two special unitary matrices $U_{k}, \tilde{U}_{k} \in \SU{N}$, the new state space $i\eta\in\su{N}$ can be defined by Lie algebra element $\eta^\wedge_k = -i \log(U^{\dagger}_k \tilde{U}_k)$. For the special unitary group, equipped with a bi-invariant metric such as the Frobenius norm, the geodesic connecting $U_k$ and $\tilde{U}_k$ can be be parametrized by $s\in \R$ as $X_k(s) = U_k \exp(i s \eta^\wedge_k)$, which satisfies $X_k(0) = U_k \text{ and } X_k(1) = \tilde{U}_k$.
	
	Combining Eq.~\eqref{eq:su2_perturbed_algebra} with Eq.~\eqref{eq:su2_pauli_to_algebra_perturb}, allows us to express perturbations in the controls $u_k$ as perturbations of Lie algebra elements $\xi^\wedge_{\epsilon, k} = \xi^\wedge_{k} + \delta \xi^\wedge_{k}$. Together with Eqs.~\eqref{eq:su2_perturbed_group} and~\eqref{eq:eqm_rev}, these relations lead to discretized equations of motion for the new state variable $\eta^\wedge_k$. We provide the details of the calculation for $\SU{N}$ in Sec.~\ref{sec:app_eqm_lin} that lead to the equation
		\begin{align}
        \label{eq:eta_dyn_analytical}
        \eta^\wedge_{k+1} &= -i \log\bigl(
            \exp(-\xi^\wedge_k \Delta t) \exp(i\eta^\wedge_k)
            \exp(\xi^\wedge_{\epsilon, k} \Delta t )
        \bigr)
		\end{align}
    which is the $\SU{N}$ analogue of Eq.~(23) in Ref.~\cite{boutselis2021}. The linearized form of this analytic expression can be derived in two different ways: either by applying the Baker-Campbell-Hausdorff (BCH) expansion, or by performing a first-order Taylor expansion~\cite{boutselis2021}. We recall the BCH approach for the first-order terms in Eq.~\eqref{eq:app_dyn_eta_lin} in Sec.~\ref{sec:app_eqm_lin} of the appendix, which yields:
	\begin{align} \label{eq:dyn_eta_lin} 
		\eta^\wedge_{k+1} \approx
		&f_{\eta, k}\of{\eta_k, u_k} \of{\eta^\wedge_k} +
		f_{u, k}\of{\eta_k, u_k}\of{\sigma \delta u_k}\\ 
		f_{\eta, k}\of{\eta_k, u_k} 
		\label{eq:dyn_eta_eta}
		 &=\Ad_{\exp(-(i H_d u^k \sigma \Delta t))} \\
		f_{u, k}\of{\eta_k, u_k}
		\label{eq:dyn_eta_u}
		&= \Delta t H_d \dexp_{-( i H_d u^k \sigma \Delta t)}.
	\end{align}	

	The $\dexp$ map can be approximated by its series expansion in terms of $\ad$ operators as given in Eq.~\eqref{eq:dexp_series}. In Ref.~\cite{boutselis2021}, the authors use second-order terms in their code and in the derivation of the Lie-algebraic equation of motion. In Ref.~\cite{teng2022}, the authors only consider zero-order terms. Since iLQR already linearizes the system dynamics at each iteration, the $\dexp$ series truncation is a secondary approximation that catches some higher-order terms but not all. Beyond a sufficient order, improvements in Jacobian accuracy are expected to yield diminishing returns, as the dominant error stems from the linear approximation of the dynamics rather than the $\dexp$ truncation. In our numerical studies, we consider three orders of the $\dexp$ map series expansion. The matrix equation of motion for $\eta^\wedge_k$ in Eq.~\eqref{eq:dyn_eta_lin} enables us to express it as a matrix equation of vectors $\eta_k, \xi_k, u_k \in \R^{\dim \su{N}}$ at time step $k$ by expressing Eqs.~\eqref{eq:dyn_eta_eta} and~\eqref{eq:dyn_eta_u} as matrices on that vector space. Exploiting the commutation relations of the Pauli operators in Eq.~\eqref{eq:app_pauli_commutator} allows us
    $\of[\ad_{-i\sigma_a}]{\sigma_b} = 2f_{ab}^{\;\;\;c} \sigma_c$ to be computed outside the optimization routine, leaving only the computation of the matrix
    $[\ad_{-iH_d\Delta t u_k\sigma}\of{\sigma}]_b^{\; c} = 2H_d\Delta t u^a_k f_{ab}^{\;\;\;c}$
    by matrix-vector multiplication at each optimization step.
    We provide an explicit matrix representation for the Adjoint group representation of $\SU{2}$ in Eq.~\eqref{eq:app_Ad_su2}. For general $\SU{N}$, however, deriving a closed-form matrix representation of $\Ad_U$ is more involved. Instead, for each time step $k$, the matrix representation of $\Ad_{U_k}\of{\sigma_b} = (U_k \sigma_b U^{-1}_k)^c \sigma_c$ can be constructed by computing $[\Ad_{U_k}(\sigma)]_b^{\; c} = (U_k \sigma_b U^{-1}_k)^c$ for the matrix-vector product
    $[\Ad_{U_k}(\sigma)]_b^{\; c} \eta^b_k = (U_k \sigma_b U^{-1}_k \eta^b_k)^c$ which implies that
    \begin{align} \nonumber
		\eta_{k+1} &\approx
		[f_{\eta, k}\of{\eta_k, u_k} \of{\sigma}] \eta_k \\ \label{eq:dyn_eta_lin_vec} 
        &+ [f_{u, k}\of{\eta_k, u_k}\of{\sigma}]\delta u_k\\
        \label{eq:dyn_eta_eta_vec}
		[f_{\eta, k}\of{\eta_k, u_k} \of{\sigma}]
        &= [\Ad_{\exp(-(i H_d u^k \sigma \Delta t))} \of{\sigma}]\\
        \label{eq:dyn_eta_u_vec}
		[f_{u, k}\of{\eta_k, u_k}\of{\sigma}]
        &= \Delta t H_d [\dexp_{-( i H_d u^k \sigma \Delta t)}(\sigma)].
	\end{align}

	\subsection{Final cost in su(N)}
    The fidelity-based cost function used in Ref.~\cite{boutselis2021} is difficult to generalize to $\SU{N}$. The Lie-algebraic cost function proposed in Ref.~\cite{teng2022} is easier to generalize to larger $N$, and classical satellite examples exist in which the Lie-algebraic cost function exhibits faster convergence. Hence, we define the final cost function by the vector representation of the geodesic, $\eta_k \in \R^{\dim{\su{N}}}$, obtained via $\eta^\wedge_k = (-i\log(g_k^\dagger g_\text{goal}) )^a\sigma_a$. It serves as a measure of distance to the goal matrix and vanishes if and only if $g_T = \exp(i\theta)g_\text{goal}$ for a global phase $\theta\in\R$. This representation enables the use of standard quadratic cost functions with weighting matrices $R_u, R_f, Q_f \in \R^{\dim\su{N} \times \dim\su{N}}$ for
	\begin{alignat}{2} \nonumber
		l\of{\eta_k, u_k} &= \frac{1}{2}u^T_kR_uu_k  \text{,} \quad
		&&l_f\of{\eta_T} = \frac{1}{2}\eta_T^T Q_f \eta_T \text{,} \\
		\label{eq:running_cost_1st}
		\text{with } l_{u, k}  &= R_u u_k \text{,} \qquad &&l_{\eta,f} = -Q_f \eta_T \text{,} \\
		\label{eq:running_cost_2nd}
		l_{uu,k} &=  R_u \text{,} \qquad &&l_{\eta \eta, f} = Q_f.
	\end{alignat}
    The minus sign in $l_{\eta,f}$ comes from the chosen direction $\log(g_k^\dagger g_\text{goal})$ in the $\eta^\wedge_k$ definition.
	In principle, the running cost could also depend on $\eta_k$; however, in our simulation, we do not include these terms.

    \subsection{Restriction of the control space}
	Quantum gate synthesis typically considers a restricted set of drive terms $\mathcal{H} \subset \mathcal{P}$ with cardinality $\abs{\mathcal{H}} = K$, reflecting the hardware constraints imposed on the available control Hamiltonians. The approximation of $\dexp$ given by Eq.~\eqref{eq:dexp_series} requires computing the matrix representation of the adjoint operator $\ad$. Achieving higher accuracy demands retaining higher-order terms in this series, which involves repeated matrix multiplications. These multiplications are only well-defined for square matrices, which necessitates constructing the adjoint representation over the full basis $\mathcal{P}$. Therefore, we first compute $[\ad_{-iH_d\Delta t u_k\sigma}\of{\sigma}]_b^{\; c}$ for all $\sigma \in \mathcal{P}$, determine $\dexp$, and only afterward select the rows corresponding to the control directions in $\mathcal{H}$. This yields the restricted derivatives $f_{u, k}\in \R^{K}$ required for the linearized system dynamics. The running cost and its derivatives are trivially restricted since $u_k \in \R^{K}$.

	\subsection{First-order smoothening effect}
	The spaces in Eqs.~\eqref{eq:su2_perturbed_group} and~\eqref{eq:su2_perturbed_algebra} can be combined to allow dependencies between controls and the equation of motion. In classic satellite examples, this is used to account for the system inertia, relating the control inputs to the body velocity~\cite{boutselis2021}. We can use this setting to enable smoother controls by including derivatives, as in the Euclidean formulation~\cite{heimann2025ilqr}. The new state-space is defined by $\zeta^\wedge_k = (\eta^\wedge_k, \delta \xi^\wedge_k)^T \in \g$ with
	\begin{align} \nonumber
		\zeta^\wedge_{k+1} &= f(\eta^\wedge_k, \xi^\wedge_k, u_k) = \of{f^\eta(\eta^\wedge_k, \xi^\wedge_k), f^\xi(\xi^\wedge_k, u^\wedge_k)}^T \\ \nonumber
		\eta_{k+1} &= f^\eta(\eta^\wedge_k, \xi^\wedge_k) = \text{Eq.}~\eqref{eq:eta_dyn_analytical}\\ \nonumber
		\xi_{k+1} &= f^\xi(\xi_k, u_k) = \xi_k + \Delta t u_k \\
		\label{eq:dyn_xi_xi}
		&f^\xi_\xi(\xi_k, u_k) = 1 \\
		\label{eq:dyn_xi_u}
		&f^\xi_u(\xi_k, u_k) = \Delta t.
	\end{align}

	The derivatives of the previous section build the fundamental parts for the total derivatives of the equation of motion
	\begin{align*}
		\zeta_{k+1} &\approx [f_\zeta\of{\zeta_k, u_k}]\zeta_k
        + [f_u\of{\zeta_k, u_k}]\delta u_k \\
		f_{\zeta, k} \of{\zeta_k, u_k} &=
		\begin{pmatrix}
			f^\eta_{\eta,k} & f^\xi_{\eta,k} \\
			f^\eta_{\xi,k} & f^\xi_{\xi,k} \\
		\end{pmatrix} =
		\begin{pmatrix}
			\eqref{eq:dyn_eta_eta_vec} & \Od \\
			\eqref{eq:dyn_eta_u_vec} & \eqref{eq:dyn_xi_xi}
		\end{pmatrix} \\
		f_{u,k} \of{\zeta_k, u_k} &=
		\begin{pmatrix} f^\eta_{u,k} \\ f^\xi_{u,k} \end{pmatrix} =
		\begin{pmatrix} \Od \\ \eqref{eq:dyn_xi_u} \end{pmatrix}
	\end{align*}
	and cost function
	\begin{align*}
		l\of{\zeta_k, u_k} &= \xi^T_k R_\xi \xi_k + u_k^T R_u u_k \\
		l_f\of{\zeta_T} &= \eta_T^T Q_\eta \eta_T + \xi^T_T Q_\xi \xi_T
	\end{align*}
	with trivial derivatives including Eqs.~\eqref{eq:running_cost_1st} and \eqref{eq:running_cost_2nd}.

	\subsection{Complexity analysis}
    Matrix multiplication in $\SU{N}$, general matrix inversion, matrix exponential, and the matrix logarithm have complexity $\order{N^3}$. At the same time, multiplying matrices that act on $\eta_k\in \R^{N^2-1}$ has complexity $\order{N^6}$.
	
	\textbf{Backward pass}: The product $U_k \sigma_j$ require $\order{N^3}$, while the inverse $U^{-1}_k$ can be computed with complexity $\order{N^3}$ or less when unitary is exploited. Since the Pauli-word basis $\mathcal{P}$ forms an orthonormal basis, the coefficients of  $M^\wedge = M^j\sigma_j$ can be obtained via $M^j = \trace(\sigma_j M^\wedge) = \sum_{k,l} \sigma_{jkl}M_{lk}$ which requires $\order{N^2}$ operations. Consequently, computing $(U_k \sigma_jU^{-1}_k)^\vee$ has complexity $\order{N^4}$ for each basis element $j$. Similarly, evaluating $[\ad_{\eta_k}]_{j} = [\eta_k^\wedge, \sigma_j]^\vee$ also has complexity $\order{N^4}$ per column. The first-order derivatives $f_g$ and $f_u$ are obtained by constructing all columns $j\in\{1,\ldots, N^2-1\}$ of $[\Ad]_j$ and $[\ad]_j$. Therefore, the overall complexity of computing these Jacobians is $\order{N^6}$. The derivative of the running cost with respect to the state $l_{g,k}$ is of $\order{N^4}$ if running costs on states are enabled. Let $K \leq N^2-1$ denote the number of controls. Then, computing $l_{u,k}$ requires $\order{K}$ operations since $R_u$ is diagonal. The second derivatives satisfy  $l_{gg,k}$, $l_{uu,k}$, $l_{gu,k} = \order{1}$. Computing the optimal control update $\delta u^\star_k$ as well as the value-function updates $V_{g,k}$  and $V_{gg,k}$ involves matrix multiplications and inversions of matrices in $N^2-1$ and therefore introduces no additional complexity beyond $\order{N^6}$.
    
    \textbf{Forward pass}: Calculate the control updates $u_k^\text{new}$ has complexity $\order{N^4}$ because it requires to calculate $\eta_k$ with complexity $\order{N^4}$, matrix-vector multiplication $K_k\eta_k$ with  $\order{N^4}$ and addition. Similarly, propagating the new state $g_{k+1}^\text{new}$ requires matrix exponentials and multiplications with complexity $\order{N^3}$.
	
	Altogether, the forward and backward passes over $T$ time-steps and $K\leq N^2-1$ controls yield an overall worst-case complexity of $\order{TN^6}$ per optimization iteration.
	
	In the Euclidean iLQR version~\cite{heimann2025ilqr}, the unitary matrix is represented by a real-valued vector of dimension $2N^2$. Converting between the unitary matrix and the isomorphic presentation by storing and reading the real and imaginary parts separately has complexity $\order{N^2}$. Propagating unitary matrices via the Schrödinger equation requires $\order{N^3}$. The isomorphic equation of motion $f^{\text{iso}}$ consists of switching between the representations and applying the Schrödinger equation between the switches, which takes $\order{N^3}$ operations. Hence, using automatic differentiation to compute the Jacobian $f_x$, resp. $f_u$, increases this complexity by $\order{N^2}$ to $\order{N^5}$, resp. by $\order{K}$ to $\order{K N^3}$. Computing the locally optimal update for the second derivative of the cost-to-go function $V_{gg,k}$ requires $\order{N^6}$ operations due to the matrix multiplication in the isomorphic state space. This dominates the update calculations for the first derivative of the cost-to-go function $V_{g,k}$ and controls $\delta u^\star_{k}$. While the Euclidean iLQR formulation has several subroutines that require fewer operations, its complexity is $\order{TN^6}$, as in the Lie-group version. For low values of $N$, the constants matter more. One advantage of the Lie-group formulation is that the state space dimension is $N^2-1$ in comparison to $2N^2$ for the Euclidean variant. This fact can lead to faster wall-clock runtime for the Lie-group iLQR formulation, especially for small $N$.

    The computational complexity cost of GRAPE is $\order{TKN^3}$ and of GEOPE is $\order{TKN^4}$~\cite{lewis2025}. Both variants benefit from the fact that a smaller $K$ directly decreases the computational complexity, in contrast to both iLQR variants, which are dominated by matrix multiplication in the state space of $\order{N^2}$ regardless of the number of accessible controls $K$.
	
	\section{Numerical results}
	\label{sec:results}

We implement the Euclidean and the Lie-group iLQR versions using the Python library \texttt{jax}~\cite{jax2018github}. The corresponding code will be publicly available in Ref.~\cite{heimann2026LieILQRcode} upon acceptance of this manuscript. Matrix exponentials are evaluated using \texttt{jax.scipy.linalg.expm}. Since \texttt{jax} does not provide a native implementation of the matrix logarithm, we compute it via eigenvalue decomposition using \texttt{jax.numpy.linalg.eig}. Preliminary simulations showed that simple Taylor and Pad\'e approximations do not provide sufficient numerical accuracy and stability for the optimization procedure. As mentioned in Sec.~\ref{sec:eqm}, for the approximation of the $\dexp$ map in Eq.~\eqref{eq:dexp_series}, we retain terms up to third order in the series expansion.

On current quantum hardware of superconducting qubits, pulses are provided for a native gate set, e.g., the cross-resonance gate for fixed-frequency coupled transmons~\cite{chow2011}, controlled-$\X$ gate for inductively coupled fluxoniums~\cite{lin2025}, controlled-$\Z$ gates for tunable-couplers~\cite{marxer2025} and several others. Gates that are not part of the native gate set are typically transpiled into circuits composed of native gates. Consequently, implementing such gates directly by providing the corresponding pulse envelopes can reduce both the accumulated gate error and the execution time on NISQ devices~\cite{egger2023}. In fault-tolerant quantum computing, quantum error-correcting codes are essential, and parity-check circuits constitute fundamental building blocks. These circuits therefore represent natural candidates for direct pulse-level implementations~\cite{uestuen2024}. Therefore, we test the $\CCZ$, Fredkin, and Toffoli gates as example three-qubit gates that are typically not included in the native gate set. In addition, we consider $\X$- and $\Z$-parity-check gates, $\WX_p$ and $\WZ_p$, which can be implemented as $(p+1)$-qubit gates using an auxiliary qubit~\cite{lewis2024}. The corresponding matrices are provided in Eqs.~\eqref{eq:gates_1st}--\eqref{eq:gates_last} in the appendix.

\subsection{Full Hamiltonian examples}
In this section, all elements of the algebra basis $\mathcal{P}$, constructed from Kronecker products of the single-qubit Pauli operators, are admissible as control Hamiltonians. Consequently, the full Lie algebra can be accessed through the drive Hamiltonian, and no constant drift Hamiltonian $H_0$ is included. This is an idealized benchmark setting that establishes an upper bound on performance.

\subsubsection{One-qubit example}
\begin{figure}[t]
    	\centering
    	\begin{subfigure}[t]{0.5\columnwidth} 
    		\centering
    		\includegraphics{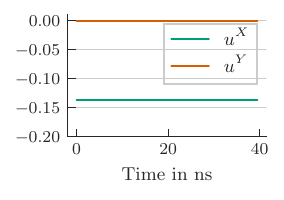}
    		\caption{Plain}
    		\label{fig:1q2l_ibm_plain}
    	\end{subfigure}%
    	~
    	\begin{subfigure}[t]{0.5\columnwidth} 
    		\centering
    		\includegraphics{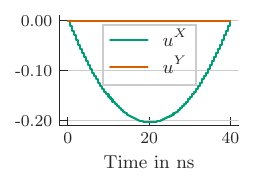}
    		\caption{Smooth}
    		\label{fig:1q2l_ibm_smooth}
    	\end{subfigure}
    	\caption{One-qubit, two-level $\X$ gate synthesized with $80$ discrete time-steps. Both Lie-group variants achieve an infidelity below $10^{-15}$ after the first optimization iteration.}
    	\label{fig:1q2l_ibm}
    \end{figure}

    We begin with the simplest case of a single qubit with two energy levels and choose $T=80$ time-steps together with $H_d = 1/(2\pi)$. Both the Lie-group iLQR formulation with and without smoothed controls achieve an infidelity below $10^{-15}$ after the first optimization iteration. The resulting control amplitudes are depicted in Fig.~\ref{fig:1q2l_ibm}; the smoothing effect introduced by the extended state-space formulation is illustrated in Fig.~\ref{fig:1q2l_ibm_smooth}. The controls obtained by the plain and the derivative Lie-group formulation implement a $-i\X$-gate, which is why the values of the $u^X$ control are negative. For this example the analytic expressions for linearized system dynamics $f_{\eta, k}$ and $f_{u, k}$ in $\SU{2}$ can be employed using Eq.~\eqref{eq:app_Ad_su2} and Eq.~\eqref{eq:app_ad_su2}, respectively. The Euclidean implementation requires three optimization steps when the target gate is modified to have unit determinant, e.g., by multiplying the $\X$-gate by $i$. In contrast, it fails to find a solution when the determinant is not equal to one, since the Euclidean cost function is not invariant to global phases.
    
    \begin{table}[b]
	\caption{Optimization iterations required to reach infidelities below $10^{-15}$ for all benchmark gates without minimal manual hyperparameter tuning and the same set across all gates for each method.}
	\begin{tabular}{lcccccc}
		\toprule
		Qubits   & 2    & 3   & 3       & 3       & 3        & 3      \\ 
		Gate     & CNOT & CCZ & Fredkin & Toffoli & $WX_2$   & $WZ_2$ \\ 
		\midrule
		Lie      &  1   &  1  &  1      &  1      &  1       &  1     \\ 
		GEOPE    &  2   &  2  &  2      &  2      &  2       &  2     \\
		Euc      & 22   & 16  & 19      & 21      & 11       & 10     \\ 
		\bottomrule
	\end{tabular} \\
	\centering
	\begin{tabular}{lcccc}
		\toprule
		Qubits   & 4      & 4      & 5      & 5      \\ 
		Gate     & $WX_3$ & $WZ_3$ & $WX_4$ & $WZ_4$ \\ 
		\midrule
		Lie      &  1     &  1     &  1     &  1     \\ 
		GEOPE    &  2     &  2     &  2     &  2     \\
		Euc      & 11     & 17     & 11     & 17     \\ 
		\bottomrule
	\end{tabular}
	\label{tab:all_N2}
\end{table}

\subsubsection{Multi-qubit examples}
\label{sec:all_N1}
    Beyond $\SU{2}$, deriving closed-form matrix expressions for the adjoint representation of the group becomes substantially more challenging. Therefore, we instead compute the corresponding matrix for $f_{\eta,k}$ at each optimization iteration and at each time step by explicitly computing its columns according to Eq.~\eqref{eq:dyn_eta_eta_vec}.

    As in the previous section, we consider the single-time-step setting and allow the drive Hamiltonian to access the full algebra basis. The Euclidean formulation requires nonzero initial amplitudes of approximately $10^{-2}$ and several optimization iterations to obtain solutions with infidelity below $10^{-15}$. In contrast, the Lie-group iLQR amplitudes can be initialized with zero while still yielding solutions within a single optimization iteration. The corresponding amplitudes are depicted in Fig.~\ref{fig:all_N1_g} in the appendix. A similar behavior is observed for the time-independent GEOPE method~\cite{lewis2024}, where zero-initialized amplitudes together with access to the full algebra basis are likewise sufficient to obtain solutions with infidelities below $10^{-10}$ in a single optimization iteration and below $10^{-15}$ with one additional iteration. The corresponding results are summarized in Tab.~\ref{tab:all_N2}. The results are obtained with minimal manual hyperparameter tuning, applying a common set of hyperparameters across all benchmark gates for each method. For GEOPE, we have not adjusted any hyperparameters. This demonstrates that the methods are robust to hyperparameter choices and broadly applicable without gate-specific tuning in this idealized case where all algebra basis elements are available as control terms. In this setting, the Lie-group iLQR variant also converges to solutions with infidelities below $10^{-15}$ in a single optimization step if more time steps, e.g., $T=12$, are allowed.

\subsection{2-local Hamiltonian examples}
\label{sec:2local_ham}

In this section, the drive Hamiltonian is restricted to one- and two-body interaction terms~\cite{lewis2024}. This increases the complexity of the optimization problem compared with the previous subsection, necessitating hyperparameter search. All results in this section are obtained without the amplitude smoothing extension. In this section, we do not compare to the GEOPE method for time-discretized pulse envelopes~\cite{lewis2025}, as its implementation, to the best of our knowledge, is not publicly available.

\subsubsection{Hyperparameter search with Bayesian optimization}
    On one hand, the entire hyperparameter space of iLQR is too large to exhaustively optimize all relevant parameters over a broad range of values. On the other hand, individual hyperparameters have well-defined purposes, enabling the manual selection of reasonable initial values to serve as starting points for a more refined search procedure. Throughout this section, we fix the number of discrete time steps to $T=12$, and apply this choice uniformly across all qubit sizes to limit the hyperparameter search space. This choice provides a decent number of optimization variables while limiting computational effort and aligns with the Horizon used in the literature for three-qubit gates~\cite{lewis2025}. This results in three primary hyperparameters: the maximum magnitude of the initial control amplitudes $u_\text{init}$, and the diagonal entries of the running and final cost matrices $R_u$ and $Q_f$, respectively. We first manually evaluated several reasonable parameter choices and selected combinations that yielded stable optimization behavior for a few seeds across multiple gates with equal numbers of qubits. These hyperparameter sets are then used as initialization points for a Bayesian optimization procedure, in which the three hyperparameters are continuously optimized over the range $10^{-5}$ to $10^5$ around the manually selected values. For Bayesian optimization, we use the upper confidence bound acquisition function with parameters $\kappa = 5$ and $\alpha = 0.01$. These values were selected based on common choices in the literature and our own tests. We decide to take $60$ optimization points: starting with $27$ runs probing the factors $\{10^{-5},10^{0},10^{-5}\}$ for all three hyperparameters adding $3$ random probes and performing $30$ Bayesian optimization steps. For comparison, we employ the same procedure for the initial hyperparameter values and the Bayesian optimization scheme for the Euclidean iLQR variant.

	We utilize the Bayesian optimization provided in the Python package~\cite{nogueira2014} to maximize the following score in the next two subsections:
    \begin{align}\label{eq:bo_min}
    	C(p) = \operatorname{median}_{a=1}^{N_s}\F(\xi^{(a)}_T, U_g)
    \end{align}
	where $p$ denotes the hyperparameter configuration, $N_s=25$ is the number of random seeds evaluated per Bayesian optimization step, and  $\F(\xi^{(a)}_T, U_g)$ denotes the fidelity obtained from the final optimized amplitudes $\xi^{(a)}_T$ of seed $a$ with respect to the target unitary $U_g$.

    Focusing on the median fidelity allows a certain number of runs to get trapped in suboptimal minima, which would otherwise substantially distort mean-based performance measures. Considering only the amplitudes $\xi^{(a)}_T$ of the last optimization iteration emphasizes the best achievable fidelities and, hence, reduces local minima. Even if the logarithmic area under the curve is normalized by the number of optimization steps, this cost can potentially overvalue runs that decrease fast in early optimization iterations but end at suboptimal minima. We present the results based on the Bayesian optimization score in Eq.~\eqref{eq:bo_min} in Secs.~\ref{subsec:three_qubits} and~\ref{subsec:three_to_five_qubits}, and additionally, for the logarithmic area under the curve Bayesian optimization score in Sec.~\ref{subsec:bo_logsum}.

\subsubsection{Three-qubit gates}
\label{subsec:three_qubits}
    \begin{figure*}[t]
    	\newcommand{\type}{median} 
    	\newcommand{\ql}{3q2l}
    	\newcommand{\model}{2-local}
    	\newcommand{\resultpath}{figs/3q2l_2local_N13/bo_3params_min_ns25}
    	\newcommand{\runida}{g72901}
    	\newcommand{\runidb}{iso72902}
    	\centering
        \begin{subfigure}[t]{0.33\linewidth}
    		\centering
    		\includegraphics{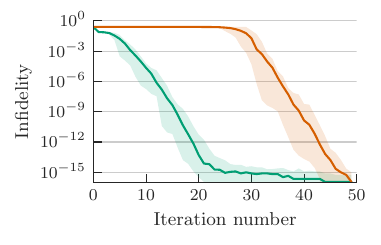}
    		\caption{CCZ}
    	\end{subfigure}%
    	~
    	\begin{subfigure}[t]{0.33\linewidth}
    		\centering
    		\includegraphics{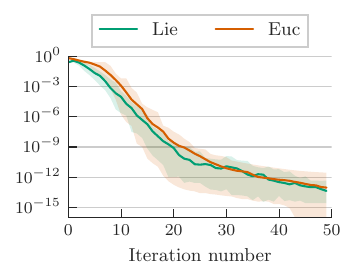}
    		\caption{Fredkin}
    	\end{subfigure}%
    	~
    	\begin{subfigure}[t]{0.33\linewidth}
    		\centering
    		\includegraphics{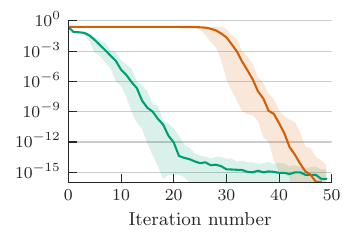}
    		\caption{Toffoli}
    	\end{subfigure}%
    	\caption{The median of 25 runs is plotted with upper and lower whiskers at each iteration step for three-qubit gates.}
    	\label{fig:3q2l_2local_N13_median_min}
    \end{figure*}

    Figure~\ref{fig:3q2l_2local_N13_median_min} shows the median infidelity over $25$ random seeds as a function of optimization iteration for both the Lie-group and Euclidean iLQR formulations applied to three-qubit benchmark gates with 2-local drive Hamiltonian terms for the best hyperparameter set obtained by Bayesian optimization with cost function Eq.~\eqref{eq:bo_min}. The shaded bands denote the lower and upper whiskers, defined as the $25\%$ and $75\%$ quartiles plus and minus $1.5$ times the interquartile range, respectively. For the CCZ and Toffoli gates, the median of the Lie-group formulation achieves lower infidelities with fewer optimization iterations. More precisely, it converges steadily from early iterations, reaching an infidelity of $10^{-15}$ by approximately iteration $25$. In contrast, the Euclidean formulation remains almost constant for the first $30$ iterations before converging rapidly to the same precision level by iteration $50$. The whiskers are pretty tight, indicating little variation across different random seeds. Plotting each run individually in Fig.~\ref{fig:3q2l_2local_N13_runs_min} in the appendix reveals that only one run for the Toffoli gate fails to find a solution (infidelity of order $10^{-1}$) whereas all other runs find solutions around infidelities of $10^{-14}$. For the Fredkin gate, the median runs of both formulations are much more similar: while the median Lie-group run reaches an infidelity of $10^{-9}$ slightly earlier, both median runs converge to a solution with infidelity $10^{-14}$.
    
    Across the three-qubit gates and both hyperparameter optimization cost functions considered, the Lie-group formulation consistently achieves lower or comparable infidelities to the Euclidean formulation, with the most pronounced advantage in the early optimization iterations.

\subsubsection{Scaling from three-qubit to five-qubit gates}
\label{subsec:three_to_five_qubits}
    \begin{figure*}[t]
    	\newcommand{\resultpath}{figs/3q2l_2local_N13/bo_3params_min_ns25}
    	\newcommand{\type}{median} %
    	\newcommand{\ql}{3q2l}
    	\newcommand{\model}{2-local}
    	\newcommand{\boconfig}{bo_3params_min_ns25}
    	\newcommand{\runida}{g72901}
    	\newcommand{\runidb}{iso72902}
    	\centering
    	\begin{subfigure}[t]{0.33\linewidth}
    		\centering
    		\includegraphics{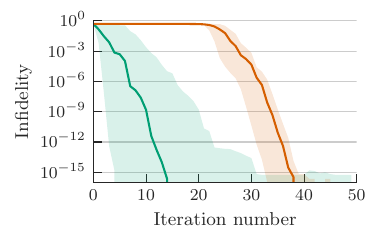}
    		\caption{$\WX_2$}
    	\end{subfigure}
    	\begin{subfigure}[t]{0.33\linewidth} 
    		\newcommand{\resultpathnew}{figs/4q2l_2local_N13/bo_3params_min_ns25}
    		\newcommand{\runidanew}{g72978}
    		\newcommand{\runidbnew}{iso72899}
    		\centering
    		\includegraphics{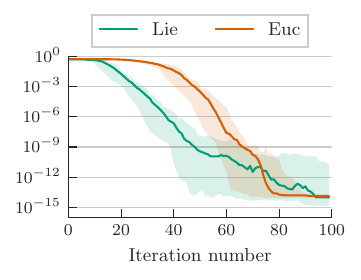}
    		\caption{$\WX_3$}
    	\end{subfigure}%
    	\begin{subfigure}[t]{0.33\linewidth} 
    		\newcommand{\resultpathnew}{figs/5q2l_2local_N13/bo_3params_min_ns25}
    		\newcommand{\runidanew}{g72980}
    		\newcommand{\runidbnew}{iso72982}
    		\centering
    		\includegraphics{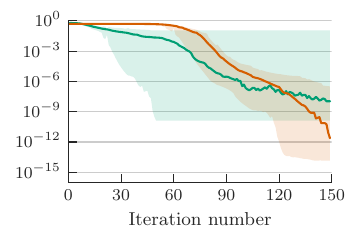}
    		\caption{$\WX_4$}
    	\end{subfigure}%
    	\\
    	\begin{subfigure}[t]{0.33\linewidth}
    		\centering
    		\includegraphics{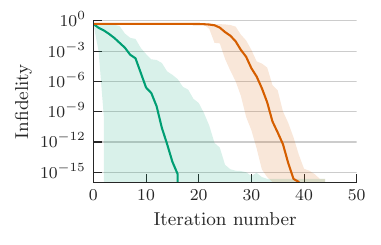}
    		\caption{$\WZ_2$}
    	\end{subfigure}%
    	\begin{subfigure}[t]{0.33\linewidth} 
    		\newcommand{\resultpathnew}{figs/4q2l_2local_N13//bo_3params_min_ns25}
    		\newcommand{\runidanew}{g72898}
    		\newcommand{\runidbnew}{iso72900}
    		\centering
    		\includegraphics{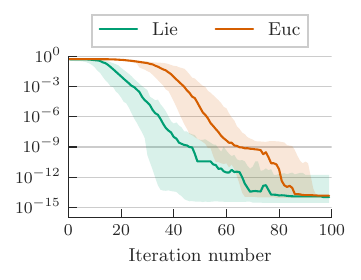}
    		\caption{$\WZ_3$}
    	\end{subfigure}%
    	\begin{subfigure}[t]{0.33\linewidth} 
    		\newcommand{\resultpathnew}{figs/5q2l_2local_N13//bo_3params_min_ns25}
    		\newcommand{\runidanew}{g72981}
    		\newcommand{\runidbnew}{iso72983}
    		\centering
    		\includegraphics{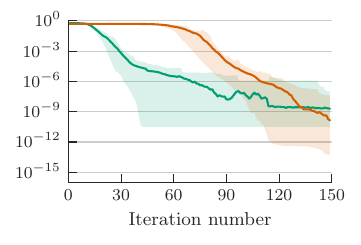}
    		\caption{$\WZ_4$}
    	\end{subfigure}%
    	\caption{The median of $25$ runs is plotted with upper and lower whiskers at each iteration step for weighted parity $\X$ and $\Z$ three-, four-, and five-qubit gates.}
    	\label{fig:wNp_2local_N13_median_min}
    \end{figure*}

    Similar results hold for the $\X$ and $\Z$ parity gates that we test for three, four, and five qubits. For the three-qubit $\WX_2$ and $\WZ_2$ gates, the faster convergence rate of the median run is most pronounced. The Lie-group formulation depends more on the initial amplitudes as represented by the broader whiskers and, as the individual runs in Fig.~\ref{fig:wNp_2local_N13_median_min} show, converges to local minima more often than the Euclidean formulation. In the cases of the four-qubit $\WX_3$ and $\WZ_3$ gates, the observed difference is comparatively smaller, though it persists notably in the early iterations of the optimization. For the five-qubit $\WX_4$ and $\WZ_4$ gates neither variant achieves infidelities below $10^{-9}$ within the $150$-iteration budget. Furthermore, while the Lie-group version reaches lower infidelities faster at the beginning of the optimization, the final infidelities are worse than those from the Euclidean variant, which continues to fall slightly until our maximum iteration of $150$, which we set to limit computational cost. As expected from results with the time-independent GEOPE~\cite{lewis2024}, both variants need more optimization iterations to lower the infidelities when the number of qubits is increased. All individual runs are depicted in Fig.~\ref{fig:wNp_2local_N13_runs_min} in the appendix.

    Scaling the $\X$ and $\Z$ parity gates from three to five qubits for 2-local drive Hamiltonian terms confirms that the Lie-group iLQR formulation can converge faster to optimal solutions than the Euclidean iLQR version. However, the example gates show a stronger tendency to local minima, and the best achieved fidelities degenerate as the number of qubits increases.

\subsubsection{Example for varying the orders of the $\dexp$ series}
\label{sec:dexp_orders}
    \begin{figure}[t]
    	\newcommand{\resultpath}{figs/dexp}
    	\newcommand{\gate}{w3px}
    	\newcommand{\runida}{g202606260}
    	\newcommand{\runidb}{g202606261}
    	\newcommand{\runidc}{g202606263}
    	\newcommand{\runidd}{g202606267}
    
    	\centering
    	\begin{subfigure}[t]{\linewidth}
    		\centering
    		\includegraphics{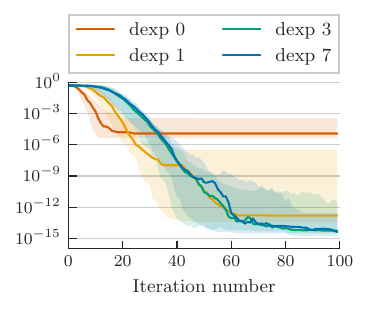}
    		\caption{Median}
            \label{fig:dexp_median}
    	\end{subfigure} \\
        \centering
    	\begin{subfigure}[t]{\linewidth}
    		\centering
    		\includegraphics{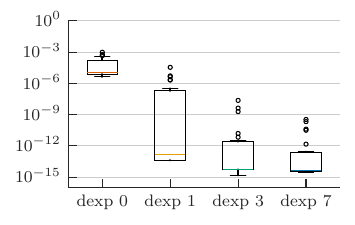}
    		\caption{Boxplot comparison of the final infidelities}
            \label{fig:dexp_box}
    	\end{subfigure}
    	\caption{Comparison of different orders in the $\dexp$ series of Eq.~\eqref{eq:dexp_series} for the $\WX_3$ four-qubit gate. We consider orders $\{0,1,3,7\}$ and perform a Bayesian optimization for each choice. The plots show the data from the 25 optimization runs using the best hyperparameter set.}
    	\label{fig:dexp}
    \end{figure}
    
    We assess the impact of varying the $\dexp$ series truncation order in Eq.~\eqref{eq:dexp_series} for the optimization of the four-qubit parity $\X$ gate, which does not achieve machine-precision infidelities. The plots in Fig.~\ref{fig:dexp} show infidelities across $25$ optimization runs, with the best hyperparameter set obtained via independent Bayesian optimization for each truncation order. The median plot in Fig.~\ref{fig:dexp_median} shows that increasing the truncation order from zero to three reduces the final infidelity, whereas increasing it further from three to seven yields no further improvements in the infidelities. The box plots of the final infidelities in Fig.~\ref{fig:dexp_box} undermine this finding and reveal that the run-to-run variance decreases slightly as the number of orders increases from three to seven. Hence, the data support the argument in Sec.~\ref{sec:eqm} that the $\dexp$ series can only improve precision up to a certain limit and motivate our choice of three orders throughout our numerical studies.    

\subsubsection{Optimization with area under the curve Bayesian optimization score}
\label{subsec:bo_logsum}
    \begin{figure*}[htbp]
    	\newcommand{\type}{median}
    	\newcommand{\ql}{3q2l}
    	\newcommand{\model}{2-local}
    	\newcommand{\boconfig}{bo_3params_logsum_ns25}
    	\newcommand{\resultpath}{figs/3q2l_2local_N13/bo_3params_logsum_ns25}
    	\newcommand{\runida}{g72911}
    	\newcommand{\runidb}{iso72912}
    	\centering
    	\begin{subfigure}[t]{0.33\linewidth} 
    		\centering
    		\includegraphics{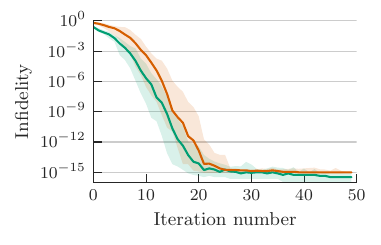}
    		\caption{CCZ}
    	\end{subfigure}%
    	~
    	\begin{subfigure}[t]{0.33\linewidth} 
    		\centering
    		\includegraphics{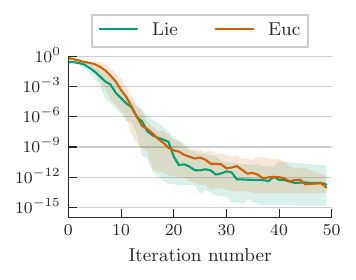}
    		\caption{Fredkin}
    	\end{subfigure}%
    	~
    	\begin{subfigure}[t]{0.33\linewidth} 
    		\centering
    		\includegraphics{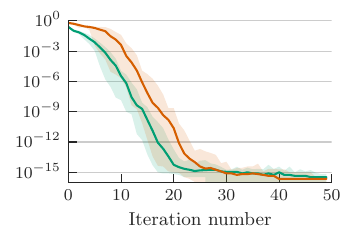}
    		\caption{Toffoli}
    	\end{subfigure}%
    	\caption{The median of $25$ runs is plotted with upper and lower whiskers at each iteration step for three-qubit gates with area under the curve Bayesian optimization cost function.}
    	\label{fig:3q2l_2local_N13_median_logsum}
    \end{figure*}

    \begin{figure*}[t] 
    	\newcommand{\resultpath}{figs/3q2l_2local_N13/bo_3params_logsum_ns25}
    	\newcommand{\type}{median} 
    	\newcommand{\ql}{3q2l}
    	\newcommand{\model}{2-local}
    	\newcommand{\boconfig}{bo_3params_logsum_ns25}
    	\newcommand{\runida}{g72911}
    	\newcommand{\runidb}{iso72912}
    	\centering
    	\begin{subfigure}[t]{0.245\linewidth} 
    		\centering
    		\includegraphics{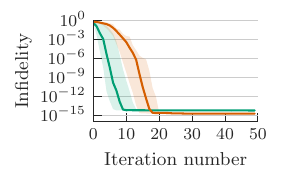}
    		\caption{$WX_2$}
    	\end{subfigure}
    	\begin{subfigure}[t]{0.245\linewidth} 
    		\centering
    		\includegraphics{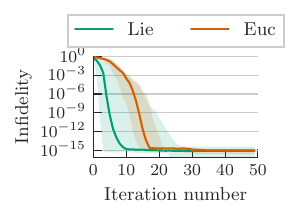}
    		\caption{$WZ_2$}
    	\end{subfigure}%
    	\begin{subfigure}[t]{0.245\linewidth} 
    		\newcommand{\resultpathnew}{figs/4q2l_2local_N13/bo_3params_logsum_ns25}
    		\newcommand{\runidanew}{g72979}
    		\newcommand{\runidbnew}{iso72923}
    		\centering
    		\includegraphics{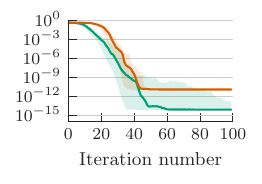}
    		\caption{$\WX_3$}
    	\end{subfigure}
    	\begin{subfigure}[t]{0.245\linewidth} 
    		\newcommand{\resultpathnew}{figs/4q2l_2local_N13/bo_3params_logsum_ns25}
    		\newcommand{\runidanew}{g72922}
    		\newcommand{\runidbnew}{iso72924}
    		\centering
    		\includegraphics{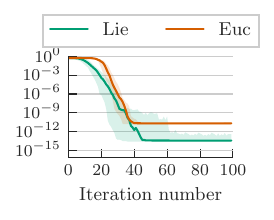}
    		\caption{$\WZ_3$}
    	\end{subfigure}
    	\caption{The median of 25 runs is plotted with upper and lower whiskers at each iteration step for parity X and Z three- and four-qubit gates with the area under the curve Bayesian optimization cost function.}
    	\label{fig:wNp_2local_N13_median_logsum}
    \end{figure*}

    We take the same initial hyperparameter sets as in Secs.~\ref{subsec:three_qubits} and~\ref{subsec:three_to_five_qubits} but perform the Bayesian optimization according to the following logarithmic area under the curve score function
    \begin{align}\label{eq:bo_logsum}
    	C(p) = \operatorname{median}_{a=1}^{N_s} \sum_{i=0}^{N_o} (-\log(1 - \F(\xi^{(a,i)}_T, U_g)))/N_o.
    \end{align}
    The logarithm of the infidelity is taken to ensure even weighting for the high-precision regime. The minus sign ensures that the cost is maximized, and the sum over all optimization iterations, $N_o$, prioritizes runs that achieve high fidelities faster. We divide the sum by the number of optimization iterations $N_o$ as a normalization factor.
    
    Figure~\ref{fig:3q2l_2local_N13_median_logsum} shows the median of $25$ runs with the best hyperparameter set with respect to that score function for three-qubit gates.  While the median for the Lie-group formulation decreases slightly earlier, the differences become less pronounced for the CCZ and Toffoli gates than they are for the minimum-based Bayesian optimization score. All individual runs are depicted in Fig.~\ref{fig:3q2l_2local_N13_runs_logsum} in the appendix.

    For the three- and four-qubit $\X$ and $\Z$ parity gates, switching to the area under the curve Bayesian optimization cost affects the steepness of the optimization improvement results less than for the other three-qubit gates of the previous section. As depicted in Fig.~\ref{fig:wNp_2local_N13_median_logsum}, the median run of the Euclidean version converges slightly faster to its optimal solution than for the minimum-based score. However, especially for the four-qubit parity gates, this comes at the cost of higher infidelities at the end of the optimization process. This can happen because the area under the curve cost in Eq.~\eqref{eq:bo_logsum} includes more information than just the precision of the last optimization iteration and, hence, balances between steepness and lowest final infidelity.  All individual runs are depicted in Fig.~\ref{fig:wNp_2local_N13_runs_logsum} in the appendix. We neglect the five-qubit case due to longer computational times and because the optimization with the minimum-based score already struggles to find high-precision solutions in this setting.
    
    While the differences between the medians of the Lie-group and the Euclidean formulation are much less pronounced in all cases, the runs for the four-qubit parity $\X$ and $\Z$ gates reveal that this cost can result in higher infidelities of final optimization results.   
    
\subsubsection{Wall-clock runtime}
    For the simulations in Secs.~\ref{subsec:three_qubits} and~\ref{subsec:three_to_five_qubits}, we record wall-clock time for gate optimization with 2-local drive Hamiltonian terms across 25 independent initializations, using the optimal hyperparameter set, on a dedicated cluster node with an AMD EPYC Genoa 9654 CPU with 96 cores and 600GB of RAM. In Tab.~\ref{tab:wall_clock}, we report per-iteration wall-clock times in the sense that we divide the total wall-clock time by the total number of optimization iterations taken, and average over gates with the same number of qubits. The Euclidean iLQR takes twice as long as the Lie-group version for the three-qubit gates, $27\%$ longer for the four-qubit gates, and is $32\%$ faster for the five-qubit gates. This confirms that the larger Euclidean state space dominates the calculation cost for small $N$, but the lower calculation cost for $f_x$ and $f_u$ can compensate for that effect for the five-qubit gates.
    
    \begin{table}[b]
		\caption{Wall-clock runtime normalized by the number of optimization iterations and averaged over gates with the same number of qubits in milliseconds. The $90\%$ confidence interval is also provided in milliseconds. Both the mean and confidence interval are calculated based on five different gates for three qubits and two gates for four and five qubits, respectively. The difference row is calculated by $(\text{Euc} - \text{Lie})/\text{Lie}$.}
		\begin{tabular}{lccc}
			\toprule
			Qubits    & 3    & 4    & 5  \\
			\midrule
			Lie       & 23ms $\pm$ 7ms & 159ms $\pm$ 81ms & 3038ms $\pm$ 113ms \\
			Euc       & 47ms $\pm$ 2ms & 203ms $\pm$ 69ms & 2065ms $\pm$ 128ms \\
			Diff      & 1.01 & 0.27  & -0.32 \\
			\bottomrule
		\end{tabular} 
		\label{tab:wall_clock}
	\end{table}

    \section{Discussion}
	\label{sec:discussion}

We demonstrate that the Lie-group formulation of iLQR can be adapted to quantum optimal control without increasing the worst-case time complexity in comparison with the Euclidean iLQR version. Our numerical results demonstrate that incorporating Lie-group structure into the iLQR framework can substantially improve convergence for quantum gate synthesis, particularly during the early stages of optimization.

The effect is especially pronounced in the idealized setting where all basis elements of the algebra can be accessed as drive Hamiltonian terms. In this setting, the Lie-theoretic formulation yields solutions with infidelities below $10^{-15}$ in the first optimization iteration, whereas the Euclidean iLQR formulation requires more optimization iterations. This behavior is on par with the time-independent GEOPE method which needs a second optimization step to reduce the infidelities from $10^{-9}$ to below $10^{-15}$. This underlines that exploiting Lie-group geometry can generally improve optimization behavior in unconstrained control settings.

In practice, quantum hardware is typically restricted to a limited set of admissible drive Hamiltonians, making it necessary to apply the optimization method to constrained sets of control operators. Simulations with a restricted set of drive Hamiltonians to one- and two-local operators confirm that the Lie-group iLQR variant consistently reduces infidelities faster, especially during the initial optimization phase. However, these results also reveal a stronger tendency to converge to local minima, reflected in a broader spread between optimization runs and in convergence plateaus observed for some gates. In particular, for the three-qubit $\CCZ$, Toffoli, $\WX_2$ and $\WZ_2$ gates, the median Lie-group run converges below $10^{-15}$ faster than the median run of the Euclidean formulation. This difference is reduced for the Fredkin gate. For the other gates, it can be reduced by choosing a different Bayesian optimization cost function which can, as a downside, lead to higher final infidelities. For the four- and five-qubit weighted $\X$ and $\Z$ parity gates, the median run of the Lie-group formulation reduces the infidelities faster but only to certain precision: For the four-qubit cases this is between $10^{-10}$ and $10^{-14}$, for the five-qubit cases between $10^{-6}$ and $10^{-8}$. Unfortunately, neither increasing the number of discrete timesteps $N$ nor increasing the terms of the $\dexp$ series necessarily leads to higher final fidelities. Hence, for the five-qubit example gates, after a certain number of optimization iterations, the Euclidean iLQR version reaches lower infidelities. However, when designing pulse amplitudes for quantum devices, hardware noise and model inaccuracies must ultimately be accounted for, for example by incorporating hardware feedback into the optimization routine~\cite{wu2018}. As a result, increasing the solution accuracy of noise-free simulations beyond a certain point does not necessarily translate into improved hardware gate fidelities. Furthermore, if higher simulation accuracy is desired, the controls obtained from the Lie-theoretic iLQR formulation can be used to warm-start the Euclidean variant after a certain number of iterations, thereby combining rapid initial convergence with potentially higher final accuracy. In addition, the Lie-group formulation is more sensitive to the initialization of the control amplitudes, leading to greater variability in the optimization process and to convergence to different local minima. Consequently, the Lie-group formulation appears to benefit more strongly from careful hyperparameter selection and initialization strategies.

More generally, increasing the number of qubits of the gate and restricting the drive Hamiltonian to 2-local terms also indirectly increases the restriction of the optimization problem, because for three, four, and five qubits, respectively, the Lie-algebra dimension is $63$ with $36$ 2-local terms, $255$ with $66$, and $1023$ with $105$. Hence, the numerical results indicate that the Lie-group formulation performs worse on restricted sets of drive terms than the Euclidean variant. For example, a minimal set of generators for the 2-qubit cross-resonance gate~\cite{chow2011}, the anisotropic Heisenberg model in Ref.~\cite{lewis2024}, as well as the Hamiltonian describing Rydberg atom arrays in Ref.~\cite{lewis2025} increase the complexity to a degree that makes it difficult to find solutions with the Lie-group iLQR formulation that we provide in this work.
The root cause of these limitations needs to be better understood in future work: including second-order dynamics approximation, i.e., switching to DDP, is known for faster convergence rates, and it could be worth investigating whether a corresponding extension for $\SU{N}$ can also improve final convergence, especially if combined with increasing the number of terms in the $\dexp$ series. Furthermore, using the Cartan-Killing form in the final cost function can provide additional information that benefits the optimization process.

Overall, the results indicate that the Lie-group formulation of iLQR constitutes a promising direction for quantum optimal control. The proposed approach improves convergence behavior by naturally respecting the geometry of $\SU{N}$. This advantage, however, comes at the cost of increased sensitivity to initialization and a stronger tendency toward local minima in restricted optimization settings.

    \section{Acknowledgments}

This work was funded by the German Ministry of Research, Technology and Space (BMFTR) in the project Data4Quantum: Datenvorverarbeitungen für Quantentechnologien under Grant No. 03DPS1134A as well as in the project QuaSa under Grant No. 13N17300 and 13N17301 administered by the VDI/VDE Innovation + Technik GmbH (VDI).
	
	\onecolumngrid
	\appendix	

\section{Lie group details}
	\label{app:group_ilqr}

	\subsection{Basic definitions}
    \label{app:group_ilqr_basics}

	A Lie group $\G$ is an abstract group which is also a smooth manifold on which the operations of group multiplication ($g \mapsto hg$, for $g,h\in\G$) and inversion ($g \mapsto g^{-1}$, for $g\in\G$) are smooth diffeomorphisms of $\G$ onto $\G$. $L_g:\G\to\G $ with $L_g(h)=hg$ denotes left-translation by $g\in\G$. Denote the vector space of smooth vector fields on $\G$ by $\Xvec(\G) \defeq \Xvec$. A vector field $X\in\Xvec$ is said to be left invariant if $X_x(g) = dL_gx$ for $x\in T_e\G$ an element of the identity tangent space of $\G$. The set of left invariant vector fields is a finite dimensional Lie algebra $\g \defeq T_e\G$, under the standard Lie-bracket operation for vector fields ($[\cdot, \cdot] \colon \Xvec \times \Xvec \to \Xvec$ such that $[X, Y](f)= \of[X]{\of[Y]{f}} - \of[Y]{\of[X]{f}}$) with its dual $\g^* = T^*_eG$. The tangent map of $L_g$ at $h\in\G$ is written as $T_hL_g = dL_g(h) \colon TG \to TG $. For $Y\in \g$ we have the following identities $X_Y = dL_gY = T_e L_g Y = gY$ and for any smooth, real function $f\colon \G \to \R$ we have $\of[df(g)]{X_Y} = \of[X_Y f]{g}$.

	Let $X, Y\in\Xvec$ be two vector fields on the Lie group $\G$ and $\nabla\colon\Xvec \times \Xvec \to \Xvec$ an affine connection. We denote the covariant derivative of $Y$ with respect to $X$ by $\nabla_X Y$.

	The Hessian operator is defined by $\hess f(g) \colon T_g\G \to T^*_g\G$ with $\hess f = \nabla^2 f = \nabla df$ and $\of[\hess(f)]{X,Y} = (\nabla_X df)(Y)$ for two vector fields $X,Y\in \Xvec$. We can rewrite the following expression as
	\begin{align*}
		XYf = X(df Y)
		= (\nabla_X df)Y + df \nabla_XY 
		&= (\nabla_X df)Y + \nabla_XY f
		= \of[\hess(f)]{X,Y} + \omega(X,Y) f \\
		\Rightarrow \of[\hess(f)]{X,Y} &= (XY - \omega(X,Y))f
	\end{align*}
	where we have introduced the connection function $\omega \colon \g \times \g \to \g$ given by $T_eL_g\omega(X,Y) = \nabla_{T_eL_g X} T_eL_g Y$~\cite{mahony2002}. Cartan-Schouten connections are proportional to the Lie bracket of their arguments and have the property that $\omega(X, Y) = \kappa [X, Y]$, where $\kappa=0$, $\kappa=1$, $\kappa=1/2$ correspond to the $(-)$, $(+)$, and $(0)$ Cartan-Schouten connection, respectively. All three have the property that the $\omega(X,Y)$ vanishes for all $X=Y$ which simplifies the Hessian equation to $\of[\hess^{(0)}(f)]{X,X} = XXf = X^* \hess^{(0)}(f) X$.

	We recall several useful identities~\cite{boutselis2021}
	\begin{align}
			\nonumber 
			\exp0 &= e \qquad \log{e} = 0 \\
			\nonumber 
			\dexp(0)\cdot\rho &=\rho
			\qquad \dlog(e)\cdot\rho =\rho \\
			\label{eq:app_Ad_Dexp}
			\dexp_{(-\rho)}(\lambda) &= \Ad_{\exp -\rho} \dexp_\rho (\lambda).
		\end{align}

	\subsection{Taylor expansion on Lie groups}
    \label{sec:app_taylor}

    With the definitions and identities from the previous section, we calculate the first and second derivatives of the smooth, real function $f(g_\epsilon)$ by following the proof of Theorem 2.5 in Ref.~\cite{michor1987}:
	\begin{align*}
		f\of{g_\epsilon} &= 
		f\of{g\exp(t \xi^\wedge)} \\
		c(s) &\defeq g\exp(s \xi^\wedge) \quad \text{and} \quad c(0) = g \\
		\gamma(s) &\defeq \exp(s \xi^\wedge) \quad \text{and} \quad \gamma(0) = e \\
 		\diff[]{}{s}f\of{g\exp(s \xi^\wedge)}
		&= 
		\diff[]{}{s} \left( f \circ L_g \circ \exp(s\xi^\wedge) \right)
		\\ &=
		d\of[f]{c(s)} d\of[L_g]{\gamma(s)}\xi^\wedge\gamma(s)
		\\ &= 
		d\of[f]{c(s)} T_{\gamma(s)}L_g T_eL_{\gamma(s)}(\xi^\wedge)
		\\ &=
		d\of[f]{c(s)} T_{e}L_{c(s)}(\xi^\wedge)
		\\ &= 
		d\of[f]{c(s)} X_{\xi^\wedge}
		\\ &= \of[X_{\xi^\wedge} f ]{c(s)} \\
		\diff[2]{}{s}f\of{g\exp(s \xi^\wedge)} &= \diff[]{}{s} \left(\of[X_{\xi^\wedge} f ]{c(s)}\right)
		= \of[X_{\xi^\wedge}(X_{\xi^\wedge} f) ]{c(s)}
		= \of[\hess^{(0)}(f(c(s)))]{X_{\xi^\wedge},X_{\xi^\wedge}}.
	\end{align*}

	The Lie-group Taylor expansion in direction $\xi^\wedge \in \g$ around the origin is given by
	\begin{align}
		\nonumber
		f\of{g_\epsilon} &= 
		f\of{g\exp(t \xi^\wedge)}
		\overset{\text{\cite[2.12.3]{varadarajan1974}}}{=}
		\sum_{i=0}^\infty\frac{t^i}{i!} f(g;\xi^{\wedge i})
		\overset{\text{\cite[2.12.2]{varadarajan1974}}}{=}
		\sum_{i=0}^\infty\frac{t^i}{i!} \diff[i]{}{s}
		f\of{g\exp(s \xi^\wedge)}\at[\Big]{s=0} \\ \nonumber
		&= f\of{g\exp(s \xi^\wedge)}\at[\Big]{s=0}
		+ t\of[X_{\xi^\wedge} f ]{c(s)}\at[\Big]{s=0}\at[\Big]{s=0}
		+ t^2\frac{1}{2} \of[\hess^{(0)}(f(c(s)))]{X_{\xi^\wedge},X_{\xi^\wedge}}\at[\Big]{s=0}
		+ \order{\abs{t}^3} \\ \nonumber
		&= \of[f]{g}
		+ t d \of[f]{g} T_eL_g \xi^\wedge
		+ t^2\frac{1}{2} \hess^{(0)}(f(g)) (T_eL_g \xi^\wedge, T_eL_g \xi^\wedge)
		+ \order{\abs{t}^3} \\ \nonumber
		&= \of[f]{g}
		+ t \left(T_eL^*_g \circ d \of[f]{g}\right) (\xi^\wedge)
		+ t^2\frac{1}{2} \left(
			T_eL^*_g \circ \hess^{(0)}(f(g)) \circ T_eL_g
		\right) (\xi^\wedge) (\xi^\wedge)
		+ \order{\norm{\xi^\wedge}^3} \\
		\label{eq:app_taylor_f_g}
		\delta f (g_\epsilon) &\defeq f(g_\epsilon) - f(g) =
		\underbrace{\left(T_eL^*_g \circ d \of[f]{g}\right)}_{\defeq f_g(g)} (\xi^\wedge) + 
		\frac{1}{2} \underbrace{\left(
			T_eL^*_g \circ \hess^{(0)}(f(g)) \circ T_eL_g
		\right)}_{\eqdef f_{gg}(g)} (\xi^\wedge) (\xi^\wedge) + \order{\norm{\xi^\wedge}^3}
	\end{align}
	where the $t$ is absorbed into the $\xi^\wedge$. If the function $f \colon \G \times \R^N \to \R$ depends on two arguments and we perturb as in Eqs.~\eqref{eq:su2_perturbed_group} and~\eqref{eq:su2_perturbed_algebra}, we obtain the Taylor expansion:
	\begin{align}
		\nonumber
		\delta f(g_\epsilon, u_\epsilon) &=
		f_g(g,u) (\xi^\wedge) + f_u(g,u) (\delta u)  \\
		\label{eq:app_taylor_f_gu}
		&\qquad +
		\frac{1}{2}\left(
			f_{gg}(g,u) (\xi^\wedge) (\xi^\wedge) + f_{ug}(g,u) (\delta u) (\xi^\wedge)
			+ f_{gu}(g,u) (\xi^\wedge) (\delta u) + f_{uu}(g,u) (\delta u) (\delta u)
		\right) + \order{\norm{(\xi^\wedge, \delta u)}^3} \\ \nonumber
		f_g(g,u) &= T_eL^*_g \circ d_g \of[f]{g,u} \\ \nonumber
		f_u(g,u) &= d_u f(g,u) \\ \nonumber
		f_{gg} (g,u) &= T_eL^*_g \circ \hess^{(0)}_g(f(g,u)) \circ T_eL_g \\ \nonumber
		f_{ug} (g,u) &= T_eL^*_g \circ d_g d_u \of[f]{g,u} \\ \nonumber
		f_{gu} (g,u) &= T_eL^*_g \circ d_u d_g \of[f]{g,u} \\ \nonumber
		f_{uu} (g,u) &= d^2_u \of[f]{g,u}
	\end{align}
	with the notation~\cite{boutselis2021}
	\begin{align}
			\label{eq:app_not}
			\order{\norm{(X,Y)}^n} \defeq \order{\norm{X}^{n_X} \norm{Y}^{n_Y}}
			\quad \text{ with } \quad n_X + n_Y = n.
		\end{align}

	\subsection{Equation of motion for su(N) elements}
	\label{sec:app_eqm_lin}

	We can combine the previously defined equations to derive
	\begin{align}
		\eta^\wedge_{k+1} &= \nonumber
		-i \log(U^{\dagger}_{k+1} U_{\epsilon,k+1}) \\
		&\overset{\eqref{eq:eqm_rev}}{=}
		-i \log(
		\exp(\xi^\wedge_k \Delta t)^\dagger \tilde{U}^{k \dagger}
		U_{\epsilon, k}\exp(\Delta t \xi^\wedge_{\epsilon, k})
		) \nonumber \\
		&\overset{\eqref{eq:su2_perturbed_group}}{=}
		-i \log(
		\exp(\xi^\wedge_k \Delta t)^\dagger U^{\dagger}_k
		U_k\exp(i\eta^\wedge_k) \exp(\Delta t \xi_{\epsilon,k})
		) \quad | \text{ } \xi^\wedge_k \text{ is skew} \nonumber\\
		\label{eq:suN_eqm_eta}
		&\overset{\eqref{eq:su2_perturbed_algebra}}{=}
		-i \log(
		\exp(-\xi^\wedge_k \Delta t) \exp(i\eta^\wedge_k)
		\exp((\xi^\wedge_k + \delta \xi^\wedge_k) \Delta t )
		).
	\end{align}
    which derives Eq.~\eqref{eq:eta_dyn_analytical}, the analogue of Eq.~(23) in Ref.~\cite{boutselis2021}.

	Furthermore, we present the details of deriving Eq.~\eqref{eq:dyn_eta_eta} using the Baker-Campbell-Hausdorff (BCH) formula. First, we need a few auxiliary equations. We recall the BCH formula
		\begin{align}
			\exp(A) \exp(B) &= \exp(\mu(A,B)) \text{ with } \nonumber \\
			\label{eq:app_bch}
			\mu(A,B) &= A + B + \frac{1}{2}\ad_A(B)
			+ \frac{1}{12}(\ad^2_A(B) + \ad^2_B(A)) 
			+ \frac{1}{24}\ad_A \ad^2_B(A) + \mathcal{O}(A^3, B^3),
		\end{align}
	and highlight that
	\begin{align}
			\label{eq:app_ad_square}
			\ad^2_{B_1 + B_2}(A) &= \ad^2_{B_1}(A) + \ad^2_{B_2}(A)
			+ \ad_{B_1}\ad_{B_2}(A) + \ad_{B_2}\ad_{B_1}(A)
		\end{align}
	which can be checked via brute force calculation. Then, we can rewrite the Lie-algebraic equation of motion Eq.~\eqref{eq:eta_dyn_analytical}
	\begin{align}
			\nonumber
				\exp(i\eta_k^\wedge)&\exp(\Delta t \xi_k^\wedge + \Delta t\delta\xi_k^\wedge)
				\eqdef \exp(i\eta_k^\wedge)\exp(B)\\
			\nonumber
				&\overset{\eqref{eq:app_bch}}{=} \exp \big( i\eta_k^\wedge + \Delta t \xi_k^\wedge + \Delta t\delta\xi_k^\wedge
				+\frac{1}{2}\ad_{i\eta_k^\wedge}(B) \\
			\nonumber
				&\qquad + \frac{1}{12}(\of[\ad^2_{i\eta_k^\wedge}]{B} + \of[\ad^2_{B}]{i\eta_k^\wedge}) 
				+ \frac{1}{24}\ad_{i\eta_k^\wedge}\of[\ad^2_{B}]{i\eta_k^\wedge} 
				+ \mathcal{O}(\eta_k^{3}, \delta\xi^3) \big) \\
			\nonumber
				&= \exp \big( \Delta t \xi + \Delta t\delta\xi_k^\wedge + \frac{1}{12}\ad^2_{i\eta_k^\wedge}(B)
				+ \frac{1}{24}\ad_{i\eta_k^\wedge}\ad^2_{B}(i\eta_k^\wedge) \\
			\nonumber
				&\qquad + i\eta_k^\wedge - \frac{1}{2}\ad_{B}(i\eta_k^\wedge)
				+ \frac{1}{12}\ad^2_{B}(i\eta_k^\wedge) 
				+ \mathcal{O}(\eta_k^{3}, \delta\xi_k^3) \big) \\
			\nonumber
				&\overset{\eqref{eq:dexp_inv_series}}{=}\exp \big( \Delta t \xi_k^\wedge + \Delta t\delta\xi_k^\wedge
				+ \frac{1}{12}\ad^2_{i\eta_k^\wedge}(B)
				+ \frac{1}{24}\ad_{i\eta_k^\wedge}\ad^2_{B}(i\eta_k^\wedge) + \dexp^{-1}_B(i\eta_k^\wedge) 
				+ \mathcal{O}(\eta_k^{3}, \delta\xi_k^3) \big)\\
			\nonumber
				&\overset{\eqref{eq:app_ad_square}}{=} \exp \big( \Delta t \xi_k^\wedge + \Delta t\delta\xi_k^\wedge + \dexp^{-1}_B(i\eta_k^\wedge)
				+ \frac{1}{12}\ad^2_{i\eta_k^\wedge}(\Delta t \xi_k^\wedge)
				+ \frac{1}{12}\ad^2_{i\eta_k^\wedge}(\Delta t\delta\xi_k^\wedge)\\
			\nonumber
				& \qquad 
				+ \frac{1}{24}\ad_{i\eta_k^\wedge}\left(
					\ad^2_{\Delta t \xi_k^\wedge}(i\eta_k^\wedge) +
					\ad^2_{\Delta t \delta\xi_k^\wedge}(i\eta_k^\wedge) +
					\ad_{\Delta t \xi_k^\wedge}\ad_{\Delta t \delta\xi_k^\wedge}(i\eta_k^\wedge) +
					\ad_{\Delta t \delta\xi_k^\wedge}\ad_{\Delta t \xi_k^\wedge}(i\eta_k^\wedge)
				\right)
				+ \mathcal{O}(\eta_k^3, \delta\xi^3) \big)\\
			\label{eq:app_exp_split_basic}
				&\overset{\eqref{eq:app_not}}{=} \exp \big( \Delta t \xi + \Delta t\delta\xi + \dexp^{-1}_B(i\eta_k^\wedge)
				+ \frac{1}{12}\ad^2_{i\eta_k^\wedge}(\Delta t \xi)
				+ \frac{1}{24}\ad_{i\eta_k^\wedge}\ad^2_{\Delta t \xi}(i\eta_k^\wedge) + \mathcal{O}(\norm{(\eta_k, \delta\xi)}^3).
			\end{align}
			Next, we split $\dexp^{-1}_B(i\eta_k^\wedge)$ into its terms $\Delta t \xi_k$ and $\Delta t \delta\xi_k$:
			\begin{align}
			\nonumber
				\dexp^{-1}_{\Delta t \xi_k^\wedge + \Delta t\delta\xi_k^\wedge}(i\eta_k^\wedge)
				&\overset{\eqref{eq:dexp_inv_series}}{=} 
				i\eta_k^\wedge - \frac{1}{2}\ad_{\Delta t \xi_k^\wedge + \Delta t\delta\xi_k^\wedge}(i\eta_k^\wedge)
				+ \frac{1}{12} \ad^2_{(\Delta t \xi_k^\wedge + \Delta t\delta\xi_k^\wedge)}(i\eta_k^\wedge) 
				+ \mathcal{O}(\eta_k^3, \delta\xi^3)\\
			\nonumber
				&\overset{\eqref{eq:app_ad_square}}{=} i\eta_k^\wedge - \frac{1}{2}\ad_{\Delta t \xi_k^\wedge}(i\eta_k^\wedge)
				+ \frac{1}{12} \ad^2_{\Delta t \xi_k^\wedge}(i\eta_k^\wedge) \\
			\nonumber
				&\qquad - \frac{1}{2}\ad_{\Delta t\delta\xi_k^\wedge}(i\eta_k^\wedge)
				+ \frac{1}{12}(
				\ad^2_{\Delta t\delta\xi_k^\wedge}(i\eta_k^\wedge)
				+ \ad_{\Delta t\delta\xi_k^\wedge}\ad_{\Delta t \xi_k^\wedge}(i\eta_k^\wedge)
				+ \ad_{\Delta t \xi_k^\wedge}\ad_{\Delta t\delta\xi_k^\wedge}(i\eta_k^\wedge)
				) 
				+ \mathcal{O}(\eta_k^3, \delta\xi_k^3)\\
			\nonumber
				&\overset{\eqref{eq:dexp_inv_series}, \eqref{eq:app_not}}{=}
				\dexp^{-1}_{\Delta t \xi_k^\wedge}(i\eta_k^\wedge)
				- \frac{1}{2}\ad_{\Delta t\delta\xi_k^\wedge}(i\eta_k^\wedge) \\
				\label{eq:app_dexpinv_split}
				&\qquad \quad + \frac{1}{12}(
				\ad_{\Delta t\delta\xi_k^\wedge}\ad_{\Delta t \xi_k^\wedge}(i\eta_k^\wedge)
				+ \ad_{\Delta t \xi_k^\wedge}\ad_{\Delta t\delta\xi_k^\wedge}(i\eta_k^\wedge)
				)+  \order{\norm{(\eta_k^\wedge, \delta\xi_k^\wedge)}^3}.
			\end{align}
			Combing Eq.~\eqref{eq:app_exp_split_basic} and Eq.~\eqref{eq:app_dexpinv_split} leads to
			\begin{align}
				\nonumber
				\exp(i\eta_k^\wedge)\exp(\Delta t \xi_k^\wedge + \Delta t\delta\xi_k^\wedge)
				&= \exp(i\eta_k^\wedge)\exp \bigg(
					\Delta t \xi_k + \Delta t\delta\xi_k
					+ \dexp^{-1}_{\Delta t \xi_k}(i\eta_k^\wedge)
					- \frac{1}{2}\ad_{\Delta t\delta\xi_k}(i\eta_k^\wedge)\\ \nonumber
				&\qquad \quad
				+ \frac{1}{12}\left(
				\ad_{\Delta t\delta\xi_k^\wedge}\ad_{\Delta t \xi_k^\wedge}(i\eta_k^\wedge)
				+ \ad_{\Delta t \xi_k^\wedge}\ad_{\Delta t\delta\xi_k^\wedge}(i\eta_k^\wedge)
				\right) \\ \nonumber
				&\qquad \quad + \frac{1}{12}\ad^2_{i\eta_k^\wedge}(\Delta t \xi_k)
				+ \frac{1}{24}\ad_{i\eta_k^\wedge}\ad^2_{\Delta t \xi_k}(i\eta_k^\wedge)
				+  \order{\norm{(\eta_k^\wedge, \delta\xi_k^\wedge)}^3}\bigg) \\ \label{eq:app_exp_Z}
				&\eqdef \exp(i\eta_k^\wedge)\exp(\Delta t \xi_k + Z^\wedge_k) \\ \nonumber
				Z^\wedge_k &\defeq \Delta t\delta\xi_k^\wedge + \dexp^{-1}_{\Delta t \xi_k^\wedge}(i\eta_k^\wedge)
				- \frac{1}{2}\ad_{\Delta t\delta\xi_k^\wedge}(i\eta_k^\wedge) + 
				\frac{1}{12}\big( \\ \nonumber
					&\qquad \quad
					\ad^2_{i\eta_k^\wedge}(\Delta t \xi_k^\wedge) + 
					\frac{1}{2}\ad_{i\eta_k^\wedge}\ad^2_{\Delta t \xi_k^\wedge}(i\eta_k^\wedge) +
					\ad_{\Delta t\delta\xi_k^\wedge}\ad_{\Delta t \xi_k^\wedge}(i\eta_k^\wedge) +
					\ad_{\Delta t \xi_k^\wedge}\ad_{\Delta t\delta\xi_k^\wedge}(i\eta_k^\wedge) \\ \label{eq:app_Z}
				&\quad \quad \big) + \order{\norm{(\eta_k^\wedge, \delta\xi_k^\wedge)}^3}
			\end{align}
			which is the $\SU{N}$ analogue for Eq.~(25) in Ref.~\cite{boutselis2021}. For iLQR only first-order terms are needed and, hence, we will use the first-order approximation of $Z^\wedge_k =  \Delta t\delta\xi_k^\wedge + \dexp^{-1}_{\Delta t \xi_k^\wedge}(i\eta_k^\wedge) + \order{\norm{(\eta_k^\wedge, \delta\xi_k^\wedge)}^2}$

		We need one more auxiliary equation~\cite[Eq.~(44)]{boutselis2021} which can be derived by applying the BCH formula:
			\begin{align}
				\nonumber
				\exp(-A) \exp(A+B) &= \exp \big(
				-A + A + B + \frac{1}{2}\ad_{-A}(A+B)
				+ \frac{1}{12}\left(
				\ad^2_{-A}(A+B) + \ad^2_{A+B}(-A)
				\right) \\ \nonumber
				&\quad + \frac{1}{24}\ad_{-A}\ad^2_{A+B}(-A)
				+ \order{\norm{(A,B)}^5} \big)\\ \nonumber
				&= \exp \big(B + \frac{1}{2}\ad_{-A}(B) + \frac{1}{12}\ad^2_{-A}(B) + 
				\frac{1}{12}\ad_{A+B}\ad_{A+B}(-A) \\ \nonumber
				&\quad +
				\frac{1}{24}\ad_{-A}\ad_{A+B}\ad_{A+B}(-A) + \order{\norm{(A,B)}^5} \big) \\
				\nonumber
				&= \exp \big(B + \frac{1}{2}\ad_{-A}(B) + \frac{1}{12}\ad^2_{-A}(B)
				+ \frac{1}{12} (\ad_{A}\ad_{B}(-A) + \ad_{B}\ad_{B}(-A))  \\ \nonumber
				&\quad + \frac{1}{24} (
				\ad_{-A}\ad_{A}\ad_{B}(-A) + \ad_{-A}\ad_{B}\ad_{B}(-A)
				) + \order{\norm{(A,B)}^5} \big) \\ \nonumber
				&= \exp \big(B + \frac{1}{2}\ad_{-A}(B) + \frac{1}{6}\ad^2_{-A}(B)
				+ \frac{1}{24} \ad^3_{-A}(B)\\ \nonumber
				&\quad +
				\frac{1}{12}\ad^2_{B}(-A) + \frac{1}{24}\ad_{-A}\ad^2_{B}(-A)
				+ \order{\norm{(A,B)}^5} \big) \\ \label{eq:app_exp_dexp_BCH}
				&= \exp \big(\dexp_{-A}(B) +
				\frac{1}{12}\ad^2_{B}(-A) + \frac{1}{24}\ad_{-A}\ad^2_{B}(-A)
				+ \order{\norm{(A,B)}^5} \big).
			\end{align}

	Combing previous equations reproduces Eq.~\eqref{eq:dyn_eta_eta} and Eq.~\eqref{eq:dyn_eta_u}:
	\begin{align} \nonumber
		\eta_{k+1}^\wedge &\overset{\eqref{eq:suN_eqm_eta}}{=} -i \log(
		\exp(-\Delta t\xi_k^\wedge) \exp(i\eta_k^\wedge)
		\exp(\Delta t \xi_k^\wedge + \Delta t\delta \xi_k^\wedge )
		) \\ \nonumber
		&\overset{\eqref{eq:app_exp_Z}}{=} -i \log(
		\exp(-\Delta t\xi_k^\wedge)
		\exp(
			\Delta t\xi_k^\wedge + Z^\wedge_k
		) ) \\ \nonumber
		&\overset{\eqref{eq:app_exp_dexp_BCH}}{=} -i \log( \exp(
		\dexp_{(-\Delta t \xi_k^\wedge)}(
			Z^\wedge_k
		) + \order{\norm{Z^\wedge_k}^2}
		)) \\ \nonumber
		&= -i \dexp_{(-\Delta t \xi_k^\wedge)}(Z^\wedge_k) + \order{\norm{Z^\wedge_k}^2} \\
		 \nonumber
		&\overset{\eqref{eq:app_Z}}{=} -i \left(
			\dexp_{(-\Delta t \xi_k^\wedge)}(\Delta t \delta \xi^k) +
			\dexp_{(-\Delta t \xi_k^\wedge)}(\dexp^{-1}_{\Delta t \xi}(i\eta^k))
		\right) + \order{\norm{(\eta_k^\wedge, \delta\xi_k^\wedge)}^2} \\ \nonumber
		&\overset{\eqref{eq:app_Ad_Dexp}}{=} -i \left(
			\Delta t \dexp_{(-\Delta t \xi_k^\wedge)}(\delta \xi^k) +
			\Ad_{\exp(-\Delta t \xi_k^\wedge)}
			\dexp_{(\Delta t \xi_k^\wedge}(\dexp^{-1}_{\Delta t \xi}(i\eta^k))
		\right) +\order{\norm{(\eta_k^\wedge, \delta\xi_k^\wedge)}^2}  \\ \nonumber
		&= -i \left(
			\Delta t \dexp_{(-\Delta t \xi_k^\wedge)}(\delta \xi^k) +
			\Ad_{\exp(-\Delta t \xi_k^\wedge)}(i\eta^k)
		\right) \\
		&= \Ad_{\exp(-\Delta t \xi_k^\wedge)}(\eta^\wedge_k)
			-i \Delta t \dexp_{(-\Delta t \xi_k^\wedge)}(\delta \xi_k^\wedge)
			+ \order{\norm{(\eta_k^\wedge, \delta\xi_k^\wedge)}^2} \\
		\label{eq:app_dyn_eta_lin}
		&\overset{\eqref{eq:su2_pauli_to_algebra_perturb}}{=} \Ad_{\exp(-i (H_0 + H_d u_k)\sigma \Delta t)}(\eta_k^\wedge) + 
		 \Delta t H_d \dexp_{-i (H_0 + H_d u_k)\sigma \Delta t}
		(\sigma \delta u_k) + \order{\norm{(\eta_k^\wedge, \sigma \delta u_k)}^2}
	\end{align}
    which is the $\su{N}$ analogue for the first-order terms in Eq.~(27) of Ref.~\cite{boutselis2021}.

\section{Example $\SU{2}$}\label{app:su2_basics}

	The Pauli basis is given by 
	\begin{align}\label{eq:app_pauli_basis}
		\sigma_X &=
		\begin{pmatrix}
			0 & 1 \\
			1 & 0
		\end{pmatrix} \quad \sigma_Y =
		\begin{pmatrix}
			0 & -i \\
			i & 0
		\end{pmatrix} \quad \sigma_Z =
		\begin{pmatrix}
			1 & 0 \\
			0 & -1
		\end{pmatrix} \\ \label{eq:app_pauli_commutator}
		\text{with} \quad 	&[\sigma_k, \sigma_l] = 2i\epsilon_{kl}^{\;\;\; m}\sigma_m.
	\end{align}
	
	A basis of $\su{2}$ with hermitian matrices is given by
	\begin{align*}
		E_1 = i\sigma_X &= 
		\begin{pmatrix}
			0 & i \\
			i & 0
		\end{pmatrix} \quad E_2 = i\sigma_Y = 
		\begin{pmatrix}
			0 & 1 \\
			-1 & 0
		\end{pmatrix} \quad E_3 = i\sigma_Z = 
		\begin{pmatrix}
			i & 0 \\
			0 & -i
		\end{pmatrix}\\
		\text{and}\; \quad
        &[E_k, E_l] = - [\sigma_k, \sigma_l] = -2i\epsilon_{kl}^{\;\;\; m}\sigma_m
        = -2\epsilon_{kl}^{\;\;\; m}E_m.
	\end{align*}
	In the $N=2$ case, this also forms a basis for the Lie group $\SU{2}$. Given this basis, general group element $U \in \SU{2}$ and algebra element $\xi^\wedge\in\su{2}$
	can be expressed in the Pauli basis given in Eq.~\eqref{eq:app_pauli_basis}
	\begin{align}
		\label{eq:app_su2_group_element}
		U &=
		\begin{pmatrix}
			\alpha & -\bar{\beta} \\
			\beta & \tilde{\alpha}
		\end{pmatrix}
		\quad \text{ for } \alpha, \beta \in \mathbb{C}
		\text{ with } \abs{\alpha}^2 + \abs{\beta}^2 = 1 \text{ and}\\
		\label{eq:su2_algebra_element}
		\xi^\wedge &= 
		\begin{pmatrix}
			i\xi_3  & \xi_2 + i\xi_1 \\
			-\xi_2 + i\xi_1 &  -i\xi_3
		\end{pmatrix}
		=	i\xi\sigma
		\quad \text{ for } \xi\in \R^3.
	\end{align}
	
	The hat and vee isomorphisms for SU(2) are
	\begin{alignat*}{2}
		(\cdot)^\wedge &\colon \mathbb{R}^{\operatorname{dim}(\g)} \to \g 
		\quad \text{ with }\quad
		&&\begin{pmatrix}
			\xi_1 \\ \xi_2 \\ \xi_3
		\end{pmatrix}
		\mapsto
		\begin{pmatrix}
			i\xi_3  & \xi_2 + i\xi_1 \\
			-\xi_2 + i\xi_1 &  -i\xi_3
		\end{pmatrix} = i
        \begin{pmatrix}
			\xi_3  & \xi_1 - i\xi_2\\
			\xi_1 +i\xi_2 &  -\xi_3
		\end{pmatrix} \quad \text{ and}
		\\
		(\cdot)^\vee &\colon \g \to \mathbb{R}^{\operatorname{dim}(\g)}
		\quad \text{ with }\quad 
		&&\begin{pmatrix}
			i\xi_3  & \xi_2 + i\xi_1 \\
			-\xi_2 + i\xi_1 &  -i\xi_3
		\end{pmatrix}
		\mapsto
		\begin{pmatrix}
			\xi_1 \\ \xi_2 \\ \xi_3
		\end{pmatrix}.
	\end{alignat*}
	
	Let $U\in \SU{2}$ by the basis in Eq.~\eqref{eq:app_su2_group_element}. The adjoint representation of the group, applied on elements $\eta\wedge = \eta \sigma$ is represented by the matrix:
    \begin{align}
    \label{eq:app_Ad_su2}
        [\Ad_U]_m &=
		\begin{pmatrix}
			1-2(\alpha^I \alpha^I + \beta^R\beta^R) &
			2(\alpha^R\alpha^I - \beta^R\beta^I) &
			2(\alpha^R\beta^R + \alpha^I \beta^I) \\
			-2(\alpha^R\alpha^I + \beta^R\beta^I) &
			1-2(\alpha^I\alpha^I + \beta^I\beta^I) &
			2(\alpha^R\beta^I - \alpha^I\beta^R) \\
			-2(\alpha^R\beta^R - \alpha^I\beta^I) &
			-2(\alpha^R\beta^I + \alpha^I\beta^R) &
			1-2(\beta^R\beta^R + \beta^I\beta^I)
		\end{pmatrix}
    \end{align}
    because
			\begin{align*}
				\Ad_U\sigma_X &= U\sigma_XU^\dagger 
				= \begin{pmatrix}
					\alpha & -\bar{\beta} \\
					\beta & \bar{\alpha}
				\end{pmatrix}
				\begin{pmatrix}
					0 & 1 \\
					1 & 0
				\end{pmatrix}
				\begin{pmatrix}
					\bar{\alpha} & \bar{\beta} \\
					-\beta & \alpha
				\end{pmatrix}
				= \begin{pmatrix}
					-\alpha\beta - \bar{\alpha}\bar{\beta} &
					\alpha\alpha - \bar{\beta}\bar{\beta} \\
					\bar{\alpha}\bar{\alpha} - \beta\beta &
					\alpha\beta + \bar{\alpha}\bar{\beta}
				\end{pmatrix} \\
				\Rightarrow
				x_3 &= -\alpha\beta - \bar{\alpha}\bar{\beta} = -2(\alpha^R\beta^R - \alpha^I\beta^I)\\
				x_1 + ix_2 &= \bar{\alpha}\bar{\alpha} - \beta\beta
				= 
				1-2(\alpha^I \alpha^I + \beta^R\beta^R)
				-i2(\alpha^R\alpha^I + \beta^R\beta^I)
				\\
				\Ad_U\sigma_Y &= g\sigma_Yg^\dagger
				=\begin{pmatrix}
					\alpha & -\bar{\beta} \\
					\beta & \bar{\alpha}
				\end{pmatrix}
				\begin{pmatrix}
					0 & -i \\
					i & 0
				\end{pmatrix}
				\begin{pmatrix}
					\bar{\alpha} & \bar{\beta} \\
					-\beta & \alpha
				\end{pmatrix}
				= i\begin{pmatrix}
					\alpha\beta - \bar{\alpha}\bar{\beta} &
					-\alpha\alpha - \bar{\beta}\bar{\beta} \\
					\bar{\alpha}\bar{\alpha} + \beta\beta &
					-\alpha\beta + \bar{\alpha}\bar{\beta}
				\end{pmatrix} \\
				\Rightarrow
				x_3 &= i(\alpha\beta - \bar{\alpha}\bar{\beta}) = -2(\alpha^R\beta^I + \alpha^I\beta^R)\\
				x_1 + ix_2 &= i(\bar{\alpha}\bar{\alpha} + \beta\beta)
				=
				i(1-2(\alpha^I\alpha^I + \beta^I\beta^I))
				+ 2(\alpha^R\alpha^I - \beta^R\beta^I)
				\\
				\Ad_U\sigma_Z &= g\sigma_Zg^\dagger
				\begin{pmatrix}
					\alpha & -\bar{\beta} \\
					\beta & \bar{\alpha}
				\end{pmatrix}
				\begin{pmatrix}
					1 & 0\\
					0 & -1
				\end{pmatrix}
				\begin{pmatrix}
					\bar{\alpha} & \bar{\beta} \\
					-\beta & \alpha
				\end{pmatrix}
				= \begin{pmatrix}
					\alpha\bar{\alpha} - \beta\bar{\beta} &
					2\alpha\bar{\beta}  \\
					2\bar{\alpha}\beta &
					(-\alpha\bar{\alpha} + \beta\bar{\beta})
				\end{pmatrix} \\
				\Rightarrow
				x_3 &= \alpha\bar{\alpha} - \beta\bar{\beta} = 1-2(\beta^R\beta^R + \beta^I\beta^I)\\
				x_1 + ix_2 &= 2\bar{\alpha}\beta
				=
				2(\alpha^R\beta^R + \alpha^I \beta^I) +
				i2(\alpha^R\beta^I - \alpha^I\beta^R).
			\end{align*}
		For $\zeta^\wedge \in \su{2}$ the adjoint representation of the algebra, $\ad_{\zeta^\wedge}$, is given by
			\begin{align}\nonumber
				\ad_{\zeta^\wedge}(\eta^\wedge) & =[\zeta^\wedge,\eta^\wedge]
				\overset{\eqref{eq:app_pauli_commutator}}{=} i 2\zeta^a \eta^b \epsilon_{ab}^{\;\;\; c} \sigma_c \\
                \label{eq:app_ad_su2}
				[\ad_\zeta]_m &= 2i[\zeta^a \epsilon_{ab}^{\;\;\; c}]_m
				= 2 i \begin{pmatrix}
					0 & -\zeta^3 & \zeta^2 \\
					\zeta^3 & 0 & -\zeta^1 \\
					-\zeta^2 & \zeta^1 & 0 
				\end{pmatrix}.
			\end{align}

\section{Quantum gates}
	In our numerical studies, we use the following two-level quantum gates:
	\begin{align}
		\label{eq:gates_1st}
		\CCZ &= \sum_{i,j,k={0,1}}\ket{ijk}\bra{ijk} - 2\ket{111}\bra{111} \\
		\operatorname{Fredkin} &= \sum_{i,j,k={0,1}}\ket{ijk}\bra{ijk} 
		- \ket{101}\bra{101} -\ket{110}\bra{110} + \ket{101}\bra{110} + \ket{101}\bra{110}\\
		\operatorname{Toffoli} &= \sum_{i,j,k={0,1}}\ket{ijk}\bra{ijk} 
		-\ket{110}\bra{110} - \ket{111}\bra{111} + \ket{101}\bra{111} + \ket{111}\bra{110}\\
        \WX_p &= \frac{1}{2}\of{\sigma_I^{\otimes (p+1)} + \sigma_I^{\otimes p} \sigma_X + \sigma_X^{\otimes p} \sigma_I - \sigma_X^{\otimes (p+1)}} \\
        \WZ_p &= \frac{1}{2}\of{\sigma_I^{\otimes (p+1)} + \sigma_I^{\otimes p} \sigma_X + \sigma_Z^{\otimes p} \sigma_I - \sigma_Z^{\otimes p} \sigma_X}.
		\label{eq:gates_last}
	\end{align}
    
\section{Amplitudes for idealized full drive Hamiltonian}
\begin{figure}[htbp]
		\centering
		\begin{subfigure}{0.2\textwidth}
			\centering
			\includegraphics{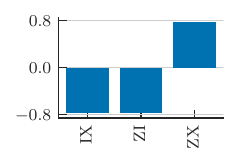}
			\caption{CNOT}
		\end{subfigure}
		\begin{subfigure}{0.2\textwidth}
			\centering
			\includegraphics{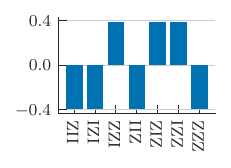}
			\caption{CCZ}
		\end{subfigure}
		\begin{subfigure}{0.2\textwidth}
			\centering
			\includegraphics{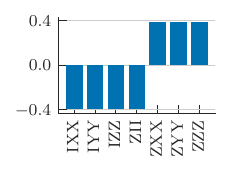}
			\caption{Fredkin}
		\end{subfigure}
		\begin{subfigure}{0.2\textwidth}
			\centering
			\includegraphics{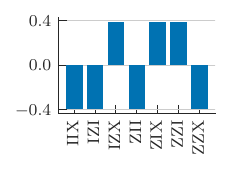}
			\caption{Toffoli}
		\end{subfigure}
		\\
		\begin{subfigure}{0.2\textwidth}
			\centering
			\includegraphics{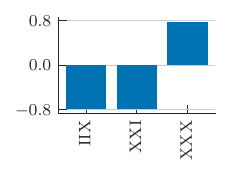}
			\caption{$WX_2$}
		\end{subfigure}
		\begin{subfigure}{0.2\textwidth}
			\centering
			\includegraphics{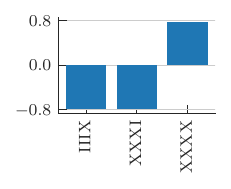}
			\caption{$WX_3$}
		\end{subfigure}
		\begin{subfigure}{0.2\textwidth}
			\centering
			\includegraphics{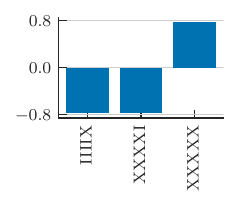}
			\caption{$WX_4$}
		\end{subfigure}
		\\
		\begin{subfigure}{0.2\textwidth}
			\centering
			\includegraphics{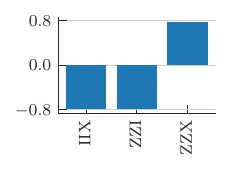}
			\caption{$WZ_2$}
		\end{subfigure}
		\begin{subfigure}{0.2\textwidth}
			\centering
			\includegraphics{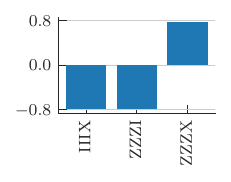}
			\caption{$WZ_3$}
		\end{subfigure}
		\begin{subfigure}{0.2\textwidth}
			\centering
			\includegraphics{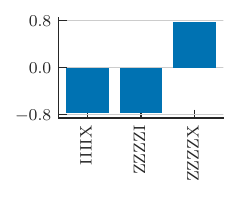}
			\caption{$WZ_4$}
		\end{subfigure}
		\caption{
			Solutions found with on iteration step with $1-\F < 10^{-15}$
			while allowing the full algebra basis as drive terms.
		}
		\label{fig:all_N1_g}
	\end{figure}

    In Sec.~\ref{sec:all_N1}, we consider the single-time-step setting and allow the drive Hamiltonian to access the full algebra basis. The Lie-group iLQR amplitudes can be initialized with zero and yield solutions with infidelities below $10^{-15}$ within a single optimization iteration.
    We provide the amplitudes in Fig.\ref{fig:all_N1_g}.

\section{Details for 2-local drive terms}
\begin{figure*}[htbp]
	\newcommand{\resultpath}{figs/3q2l_2local_N13/bo_3params_min_ns25}
	\newcommand{\type}{runs}
	\newcommand{\ql}{3q2l}
	\newcommand{\model}{2-local}
	\newcommand{\boconfig}{bo_3params_min_ns25}
	\newcommand{\runida}{g72901}
	\newcommand{\runidb}{iso72902}
	\centering
	\begin{subfigure}[t]{0.33\linewidth}
		\centering
		\includegraphics{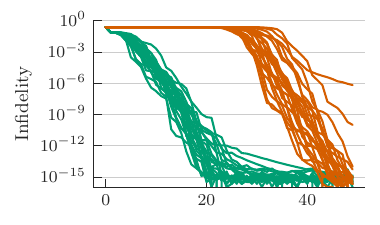}
		\caption{CCZ}
	\end{subfigure}%
	~
	\begin{subfigure}[t]{0.33\linewidth}
		\centering
		\includegraphics{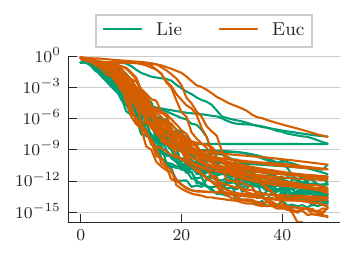}
		\caption{Fredkin}
	\end{subfigure}%
	~
	\begin{subfigure}[t]{0.33\linewidth}
		\centering
		\includegraphics{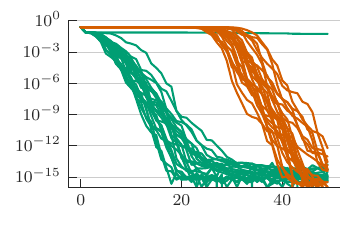}
		\caption{Toffoli}
	\end{subfigure}%
	\caption{Infidelities of the $25$ runs are plotted at each iteration step for three-qubit gates.}
	\label{fig:3q2l_2local_N13_runs_min}
\end{figure*}

\begin{figure*}[htbp]
    	\newcommand{\resultpath}{figs/3q2l_2local_N13/bo_3params_min_ns25}
    	\newcommand{\type}{runs} 
    	\newcommand{\ql}{3q2l}
    	\newcommand{\model}{2-local}
    	\newcommand{\boconfig}{bo_3params_min_ns25}
    	\newcommand{\runida}{g72901}
    	\newcommand{\runidb}{iso72902}
    	\centering
    	\begin{subfigure}[t]{0.33\linewidth}
    		\centering
    		\includegraphics{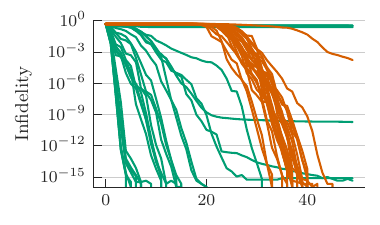}
    		\caption{$\WX_2$}
    	\end{subfigure}
    	\begin{subfigure}[t]{0.33\linewidth} 
    		\newcommand{\resultpathnew}{figs/4q2l_2local_N13/bo_3params_min_ns25}
    		\newcommand{\runidanew}{g72978}
    		\newcommand{\runidbnew}{iso72899}
    		\centering
    		\includegraphics{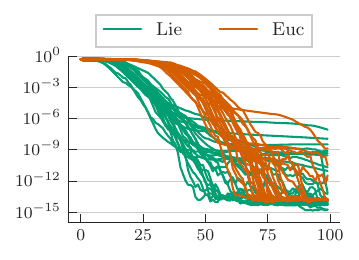}
    		\caption{$\WX_3$}
    	\end{subfigure}%
    	\begin{subfigure}[t]{0.33\linewidth} 
    		\newcommand{\resultpathnew}{figs/5q2l_2local_N13/bo_3params_min_ns25}
    		\newcommand{\runidanew}{g72980}
    		\newcommand{\runidbnew}{iso72982}
    		\centering
    		\includegraphics{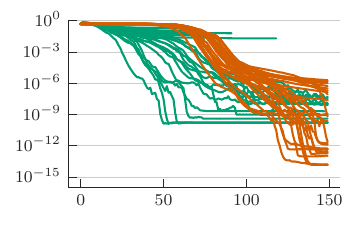}
    		\caption{$\WX_4$}
    	\end{subfigure}%
    	\\
    	\begin{subfigure}[t]{0.33\linewidth}
    		\centering
    		\includegraphics{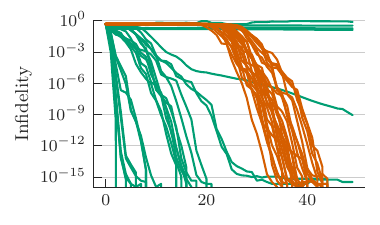}
    		\caption{$\WZ_2$}
    	\end{subfigure}%
    	\begin{subfigure}[t]{0.33\linewidth} 
    		\newcommand{\resultpathnew}{figs/4q2l_2local_N13/bo_3params_min_ns25}
    		\newcommand{\runidanew}{g72898}
    		\newcommand{\runidbnew}{iso72900}
    		\centering
    		\includegraphics{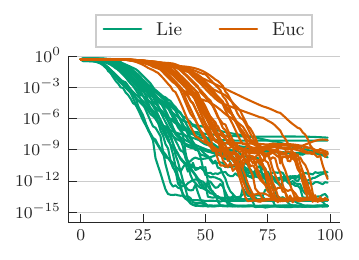}
    		\caption{$\WZ_3$}
    	\end{subfigure}%
    	\begin{subfigure}[t]{0.33\linewidth} 
    		\newcommand{\resultpathnew}{figs/5q2l_2local_N13/bo_3params_min_ns25}
    		\newcommand{\runidanew}{g72981}
    		\newcommand{\runidbnew}{iso72983}
    		\centering
    		\includegraphics{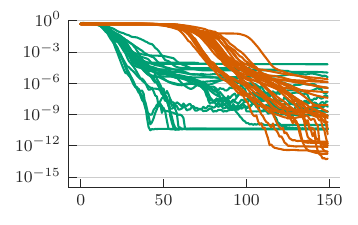}
    		\caption{$\WZ_4$}
    	\end{subfigure}%
	\caption{Infidelities of the $25$ runs are plotted at each iteration step for parity X and Z three-, four-, and five-qubit gates.}
	\label{fig:wNp_2local_N13_runs_min}
\end{figure*}

\begin{figure*}[htbp]
	\newcommand{\type}{runs}
	\newcommand{\ql}{3q2l}
	\newcommand{\model}{2-local}
	\newcommand{\boconfig}{bo_3params_logsum_ns25}
	\newcommand{\resultpath}{figs/3q2l_2local_N13/bo_3params_logsum_ns25}
	\newcommand{\runida}{g72911}
	\newcommand{\runidb}{iso72912}
	\centering
	\begin{subfigure}[t]{0.33\linewidth} 
		\centering
		\includegraphics{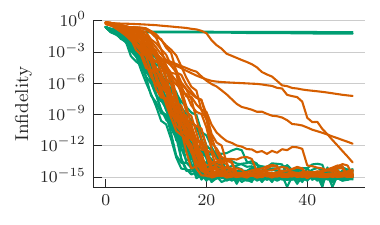}
		\caption{CCZ}
	\end{subfigure}%
	~
	\begin{subfigure}[t]{0.33\linewidth} 
		\centering
		\includegraphics{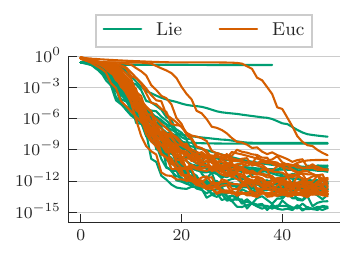}
		\caption{Fredkin}
	\end{subfigure}%
	~
	\begin{subfigure}[t]{0.33\linewidth} 
		\centering
		\includegraphics{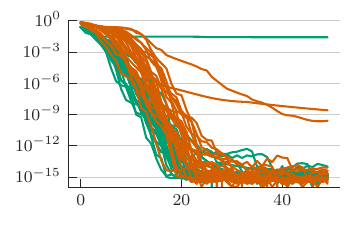}
		\caption{Toffoli}
	\end{subfigure}%
	\caption{Infidelities of the $25$ runs are plotted at each iteration step for three-qubit gates with area under the curve Bayesian optimization score function.}
	\label{fig:3q2l_2local_N13_runs_logsum}
\end{figure*}

\begin{figure*}[htbp]
	\newcommand{\resultpath}{figs/3q2l_2local_N13/bo_3params_logsum_ns25}
	\newcommand{\type}{runs} 
	\newcommand{\ql}{3q2l}
	\newcommand{\model}{2-local}
	\newcommand{\boconfig}{bo_3params_logsum_ns25}
	\newcommand{\runida}{g72911}
	\newcommand{\runidb}{iso72912}
	\centering
	\begin{subfigure}[t]{0.245\linewidth} 
		\centering
		\includegraphics{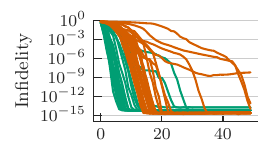}
		\caption{$WX_2$}
	\end{subfigure}
	\begin{subfigure}[t]{0.245\linewidth} 
		\centering
		\includegraphics{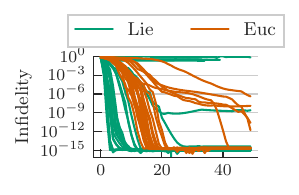}
		\caption{$WZ_2$}
	\end{subfigure}%
	\begin{subfigure}[t]{0.245\linewidth} 
		\newcommand{\resultpathnew}{figs/4q2l_2local_N13/bo_3params_logsum_ns25}
		\newcommand{\runidanew}{g72979}
		\newcommand{\runidbnew}{iso72923}
		\centering
		\includegraphics{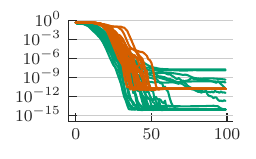}
		\caption{$\WX_3$}
	\end{subfigure}
	\begin{subfigure}[t]{0.245\linewidth} 
		\newcommand{\resultpathnew}{figs/4q2l_2local_N13/bo_3params_logsum_ns25}
		\newcommand{\runidanew}{g72922}
		\newcommand{\runidbnew}{iso72924}
		\centering
		\includegraphics{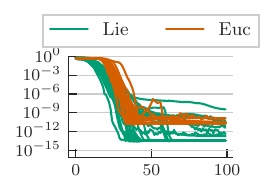}
		\caption{$\WZ_3$}
	\end{subfigure}
	\caption{Infidelities of the $25$ runs are plotted at each iteration step for parity $\X$ and $\Z$ three- and four-qubit gates with the area under the curve Bayesian optimization score function.}
	\label{fig:wNp_2local_N13_runs_logsum}
\end{figure*}

    In Sec.~\ref{sec:2local_ham}, we restrict the drive Hamiltonian to one- and two-body interaction terms and perform a Bayesian optimization. For the score function in Eq.~\eqref{eq:bo_min}, Fig.~\ref{fig:3q2l_2local_N13_runs_min} and Fig.~\ref{fig:wNp_2local_N13_runs_min} show the $25$ optimization runs individually with the best hyperparameter set for three-qubit gates and for three- to five-qubit gates, respectively.
    Fig.~\ref{fig:3q2l_2local_N13_runs_logsum} and Fig.~\ref{fig:wNp_2local_N13_runs_logsum} show the results for the same simulations expected for the Bayesian optimization score function in Eq.~\eqref{eq:bo_logsum}.
    
\newpage

    \clearpage
    \twocolumngrid
    \bibliography{main}
    
\end{document}